\documentclass[twocolumn]{autart}    

\usepackage{graphicx}          
\usepackage{epstopdf}
\usepackage{hyperref}
\hypersetup{
    colorlinks=true,
    linkcolor=blue,
    filecolor=blue,      
    urlcolor=blue,
    citecolor=cyan,
}
\newtheorem{theorem}{Theorem}[section]

\newtheorem{lemma}{Lemma}[section]

\newtheorem{remark}{Remark}[section]

\usepackage{makecell}
\usepackage{epsfig,graphicx,amssymb,amsmath}
\numberwithin{equation}{section}
\newtheorem{assumption}{\textbf{Assumption}}
\usepackage{float}
\usepackage{color}
\usepackage{xcolor}
\allowdisplaybreaks[4]
\begin{document}

\begin{frontmatter}

\title{Delayed finite-dimensional observer-based control of 2D linear parabolic PDEs} 

\thanks[footnoteinfo]{This work was supported by Israel Science Foundation (grant no. 673/19)  and by Chana and Heinrich Manderman Chair at Tel Aviv University.}

\author{Pengfei Wang}\ead{wangpengfei1156@hotmail.com},  
\author{Emilia Fridman}\ead{emilia@tauex.tau.ac.il}   

\address{School of Electrical Engineering, Tel-Aviv University, Tel-Aviv, Israel}
\begin{keyword}                           
 2D parabolic PDEs, observer-based control, time delay, vector Halanay's inequality.                                     
\end{keyword}                             

\begin{abstract}                          
Recently, a constructive method was suggested for finite-dimensional observer-based control of 1D linear heat equation, which is robust to input/output delays. 
In this paper, we aim to extend this method to the 2D case with general time-varying input/output delays (known output delay and unknown input delay) or sawtooth delays (that correspond to network-based control). 
We use the modal decomposition approach and consider boundary or non-local sensing together with non-local actuation, or Neumann actuation with non-local sensing.  
To compensate the output delay that appears in the infinite-dimensional part of the closed-loop system, for the first time for delayed PDEs we suggest a vector Lyapunov functional combined with the recently introduced vector Halanay inequality.
We provide linear matrix inequality (LMI) conditions for finding the observer dimension and upper bounds on delays that preserve the exponential stability. 
We prove that the LMIs are always feasible for large enough observer dimension and small enough upper bounds on delays. 
A numerical example demonstrates the efficiency of our method and show that 
the employment of vector Halanay's inequality allows for larger delays than the classical scalar Halanay inequality for comparatively large observer dimension.
\end{abstract}
\end{frontmatter}

\section{Introduction}\label{sec1}\vspace{-0.3cm}
Finite-dimensional observer-based controllers for PDEs are attractive in applications.
Such controllers were designed by the modal decomposition approach and have been extensively studied since the 1980s \cite{balas1988finite,christofides2012nonlinear,curtain1982finite,grune2022finite,harkort2011finite}, where efficient bound estimate on the observer and controller dimensions is a challenging problem. 
In recent paper \cite{katz2020constructive}, the first constructive LMI-based method for finite-dimensional observer-based control of 1D parabolic PDEs was suggested, where the observer dimension was found from simple LMI conditions. The results in \cite{katz2020constructive} were then extended to
input/output delay robustness in \cite{katz2021delayed,katz2022delayed,katzasampled}, delayed PDEs \cite{lhachemi2023boundary2023} and delay compensation in \cite{katz2022delayed,lhachemi2023boundary2,lhachemi2022predictor,lhachemi2023output}. 
However, the results of \cite{katz2020constructive,katz2021delayed,katz2022delayed,katzasampled,lhachemi2023boundary2,lhachemi2022predictor,lhachemi2023boundary2023,lhachemi2023output} were confined to 1D parabolic PDEs. 

\vspace{-0.15cm}
In recent years, control of high-dimensional PDEs became an active research area. 
Such systems have promising applications in engineering, water heating, metal rolling, sheet forming, medical imaging (see e.g. \cite{meurer2012control}) as well as in multi-agents deployment \cite{qi2014multi}. 
Sampled-data observers for $2$D and $N$D heat equations with globally Lipschitz nonlinearities have been suggested in \cite{am2014network,selivanov2019delayed}.  
Observer-based output-feedback controller for a linear parabolic $N$D PDEs was designed in \cite{wang2020dynamic}. In \cite{kang2021sampled}, the sampled-data control of 2D Kuramoto-Sivashinsky equation was explored. 
The results in \cite{am2014network,kang2021sampled,selivanov2019delayed,wang2020dynamic} were in rectangular domain and employed spatial decomposition approach where many sensors/actuators should be utilized. 

\vspace{-0.12cm}
 The boundary state-feedback stabilization of $N$D parabolic PDEs was studied in \cite{barbu2013boundary,munteanu2019boundary} by modal decomposition approach and in \cite{meurer2012control,liu2020boundary} by backstepping method.  
Observer-based boundary control for $N$D parabolic PDEs under boundary measurement over cubes and balls was explored in \cite{jadachowski2015backstepping,vazquez2016explicit} by the backstepping method. 
 In \cite{feng2022boundary,meng2022boundary}, observer-based control via modal decomposition approach was designed for $N$D parabolic PDEs. 
 Note that the observer designs in \cite{feng2022boundary,jadachowski2015backstepping,meng2022boundary,vazquez2016explicit} are in the from of PDEs. 
 In \cite{lhachemi2023boundary}, for the first time, the finite-dimensional observer-based control was studied for 2D and 3D parabolic PDEs under boundary actuation on an arbitrary subdomain and in-domain pointwise measurement. It was shown in \cite{lhachemi2023boundary} that the closed-loop system is stable provided the dimension of the controller is large enough. 
Note that the results in \cite{feng2022boundary,jadachowski2015backstepping,lhachemi2023boundary,meng2022boundary,vazquez2016explicit} are confined to observer-based controller design of $N$D delay-free PDEs. 
For $N$D parabolic PDEs, efficient finite-dimensional observer-based design with a quantitative bound on the observer as well as the input/output delay robustness remained open challenging problems.

\vspace{-0.12cm}
In this paper, we aim to study finite-dimensional observer-based control of linear heat equation with input/output delays in $\Omega$, an open and connected subset of $\mathbb{R}^2$. We consider either differentiable time-varying delays (unknown input delay and known output delay) or sawtooth delays (that correspond to network-based control). Based on modal decomposition approach, we consider the boundary or non-local sensing together with non-local actuation, or to Neumann actuation with non-local sensing. The novelty of this paper compared to existing results can be formulated as follows: \vspace{-0.18cm}
\begin{itemize}
   \item Compared with \cite{feng2022boundary,jadachowski2015backstepping,lhachemi2023boundary,meng2022boundary,vazquez2016explicit} for observer-based design of high-dimensional parabolic PDEs, we give efficient finite-dimensional observer-based design and provide LMI conditions for finding observer dimension and upper bounds of delays. We prove that the LMIs are always feasible for large enough observer dimension and small enough upper bounds on delays.
  \item Differently from \cite{katz2021delayed,katz2022delayed,katzasampled} for 1D parabolic PDEs where Lyapunov functional combined with classical scalar Halanay's inequality (see P. 138 in \cite{fridman2014introduction}) was suggested,  we construct vector Lyapunov functional combined with recently introduced vector Halanay's inequality (see \cite{mazenc2021vector}).  The latter allows to efficiently compensate the fast-varying output delay that appears in the infinite-dimensional part of the closed loop system essentially improving the upper bounds on delays in most of the numerical examples.
  \item Compared with spatial decomposition approach suggested in \cite{am2014network,kang2021sampled,selivanov2019delayed,wang2020dynamic} for robust stabilization of $N$D parabolic PDEs, 
  the modal decomposition approach in this paper allows for fewer actuators and sensors (including single boundary actuator or sensor).
\end{itemize}


\vspace{-0.12cm}
\textit{Notations and preliminaries:} For any bounded domain $\Omega \subset \mathbb{R}^{2}$, denote by $L^{2}(\Omega)$ the space of square integrable functions with inner product $\langle f,g \rangle=\int_{\Omega}f(x)g(x)\mathrm{d}x$ and induced norm $\|f\|_{L^{2}}^{2}=\langle f,f \rangle$.
$H^{1}(\Omega)$ is the Sobolev space of functions $f:\Omega\longrightarrow \mathbb{R}$ with a square integrable weak derivative. The norm defined in $H^{1}(\Omega)$ is $\|f\|^{2}_{H^{1}}=\|f\|^{2}_{L^{2}}+\|\nabla  f\|^{2}_{L^{2}}$, where $\nabla  f = [f_{x_{1}}, f_{x_{2}}]^{\mathrm{T}}$ and $\|\nabla  f\|^{2}_{L^{2}}=\int_{\Omega}[(f_{x_{1}})^{2}+ (f_{x_{2}})^{2}]\mathrm{d}x$.
The Euclidean norm is denoted by $|\cdot|$. For $P\in \mathbb{R}^{n\times n}$, $P>0$ means that $P$ is symmetric and positive definite. The symmetric elements of a symmetric matrix will be denoted by $*$. For $0<P\in \mathbb{R}^{n\times n}$ and  $x\in \mathbb{R}^{n}$, we write $|x|^{2}_{P}=x^{\mathrm{T}}Px$. Denote $\mathbb{N}$ by the set of positive integers.

Let $\Omega \subset \mathbb{R}^2$ be a bounded open connected set.  Following \cite{tucsnak2009observation}, we assume that either the boundary $\partial \Omega$ is of class $C^2$ or $\Omega$ is a rectangular domain. Let $\partial \Omega$ be split into two disjoint parts $\partial \Omega =\Gamma_{D}\cup \Gamma_{N}$ such that $\Gamma_{D}$ and $\Gamma_{N}$ have non-zero Lebsgue measurement. Here subscripts D and N stand for Dirichlet and for Neumann boundary conditions respectively. Let 
\vspace{-0.3cm}
\begin{equation}\label{operatorA}
	{\scriptsize \begin{array}{ll}
		\mathcal{A}\phi= -\Delta \phi, ~\mathcal{D}(\mathcal{A})= \{\phi| \phi\in H^{2}(\Omega) \cap H^{1}_{\Gamma}(\Omega)\},\\
			H^{1}_{\Gamma}(\Omega)=\{\phi\in H^{1}(\Omega)| \phi(x)=0~\mathrm{for}~ x\in\Gamma_{D}, \\
			~~~~~~~~~~~~~ \frac{\partial \phi}{\partial {\bf n}}(x)=0 ~\mathrm{for}~ x\in \Gamma_{N}\}, \vspace{-0.3cm}
	\end{array} }
\end{equation}
where $\frac{\partial}{\partial {\bf n}}$ is the normal derivative.
It follows from \cite[Proposition 3.2.12]{tucsnak2009observation} that the eigenvalues $\{\lambda_{n}\}_{n=1}^{\infty}$ of $\mathcal{A}$ are real and we can repeat each eigenvalue according to its finite multiplicity to get \vspace{-0.3cm}
\begin{equation}\label{eigenvalue2}
\begin{array}{ll}
	\lambda_{1} < \lambda_{2} \leq \dots \leq \lambda_{n} \leq \dots,~~
	\lim_{n\rightarrow\infty} \lambda_{n}=\infty.  \vspace{-0.3cm}
\end{array}
\end{equation}
We denote the corresponding eigenfunctions as $\{\phi_{n}\}_{n=1}^{\infty}$. Differently from the 1D case where $\lambda_{N}=O(N^{2})$, $N\rightarrow\infty$,  for $\lambda_{N}$, we have the following estimate which will be used for the asymptotic feasibility of LMIs: \vspace{-0.2cm}
\begin{lemma}\label{lemma}(\cite[Sec. 11.6]{strauss2007partial})
	For eigenvalues \eqref{eigenvalue2}, we have $\lim_{N\rightarrow\infty}\frac{\lambda_{N}}{N}=\frac{4\pi}{|\Omega|}$, where $|\Omega|$ is the area of $\Omega$.
\end{lemma}

\vspace{-0.2cm}
Since $\mathcal{A}$ is strictly positive and diagonalizable, we have (see Proposition 3.4.8 in \cite{tucsnak2009observation}) \vspace{-0.3cm}
\begin{equation*}
\begin{array}{ll}
	\mathcal{D}(\mathcal{A}^{\frac{1}{2}})=\{h\in L^{2}(\Omega)| \sum_{n=1}^{\infty}\lambda_{n}|\langle h,\phi_{n} \rangle |<\infty \}. \vspace{-0.3cm}
\end{array}	
\end{equation*}
Following Remark 3.4.4 in \cite{tucsnak2009observation}, we can regard $\mathcal{D}(\mathcal{A}^{\frac{1}{2}})$ as the completion of $\mathcal{D}(\mathcal{A})$ with respect to the norm $\|f\|_{\frac{1}{2}}=\sqrt{\langle \mathcal{A}f,f\rangle}=\sqrt{\sum_{n=1}^{\infty}\lambda_{n}|\langle f,\phi_{n}\rangle|^{2}}$, $f\in \mathcal{D}(\mathcal{A})$. 
For $h\in \mathcal{D}(\mathcal{A})$,  we have $\|\nabla h\|_{L^{2}}^{2} = \langle h, \mathcal{A}h  \rangle =\| h \|^{2}_{\frac{1}{2}}$, which implies \vspace{-0.3cm}
\begin{equation}\label{inequality00}
	\begin{array}{ll}
		\|\nabla h\|_{L^{2}}^{2}  = \sum_{n=1}^{\infty}\lambda_{n}h_{n}^{2}. \vspace{-0.3cm}
	\end{array} 
\end{equation}
 We have $\|f\|^{2}_{L^{2}}\leq C(\Omega)\|\nabla f\|_{L^{2}}^{2}$, $f|_{\Gamma_{D}}=0$ 
  for some constant $C(\Omega)>0$ (see \cite{2051099}), 
which together with \eqref{inequality00} implies the equivalence of $\|\cdot\|_{\frac{1}{2}}$ and $\|\cdot\|_{H^{1}}$ subject to $f(x)=0$, $x\in \Gamma_{D}$.  
We have $\mathcal{D}(\mathcal{A}^{\frac{1}{2}})=\{h\in H^{1}(\Omega) | h(x)=0, x\in \Gamma_{D}\}$. 
Finally, density of $\mathcal{D}(\mathcal{A})$ in $\mathcal{D}(\mathcal{A}^{\frac{1}{2}})$ yields that \eqref{inequality00} holds for any $h\overset{L^{2}(\Omega)}{=} \sum_{n=1}^{\infty}h_{n}\phi_{n} \in \mathcal{D}(\mathcal{A}^{\frac{1}{2}})$.

Given a positive integer $N$ and $h\in L^{2}(\Omega)$ satisfying $h\overset{L^{2}}{=}\sum_{n=1}^{\infty}h_{n}\phi_{n}$, where $h_{n}=\langle h,\phi_{n} \rangle$,  we denote $\|h\|^{2}_{N}=\sum_{n=N+1}^{\infty}h^{2}_{n}$. 
For $\phi \in L^{2}(\Omega)$ and ${\bf b}=[b_{1}, \dots, b_{d}]^{\mathrm{T}} \in (L^{2}(\Omega))^{d}$, we denote $\langle {\bf b}, \phi\rangle =[\langle b_1, \phi\rangle, \dots, \langle b_{d}, \phi \rangle ]^{\mathrm{T}}$.



\vspace{-0.2cm}
\begin{lemma}\label{lemma2}
	(Vector Halanay's Inequality \cite{mazenc2021vector}) Let $M\in \mathbb{R}^{n\times n}$ be a Metzler and Hurwitz matrix and $P\in \mathbb{R} ^{n\times n}$ be a nonnegative matrix. Let $\tau=\max\{\tau_{1},\dots,\tau_{n}\}$ with $\tau_{i}>0$ and $V=[V_{1},\dots,V_{n}]^{\mathrm{T}}: [-\tau,\infty)\rightarrow [0,\infty)^{n}$ be $C^{1}$ and \vspace{-0.3cm}
	\begin{equation*}
		\begin{array}{ll}
			\dot{V}(t)\leq MV(t)+P\sup_{s\in [t-\tau,t]}V(s), \vspace{-0.3cm}
		\end{array}
	\end{equation*}
where  $\sup_{s\in [t-\tau,t]} V(s) =\mathrm{col}\{ \sup_{s\in [t-\tau_{i},t]}V_{i}(s) \}_{i=1}^{n}$. If $M+P$ is Hurwitz, then $|V(t)|\leq D\mathrm{e}^{-\delta_{0} t}$, $t\geq 0$ for some $\delta_{0}>0$ and $D>0$.
\end{lemma}

\vspace{-0.2cm}
\section{Non-local actuation and measurement}\label{sec2}\vspace{-0.3cm}
\subsection{System under consideration and controller design}\vspace{-0.3cm}
Consider the following heat equation under delayed nonlocal actuation: \vspace{-0.3cm}
\begin{equation}\label{pde1aab}
\begin{array}{ll}
	z_{t}(x,t)=\Delta z(x,t)+qz(x,t) \\
	~~~~~~~~~~+ {\bf b}^{\mathrm{T}}(x)u(t-\tau_{u}(t)),~\mathrm{in}~ \Omega \times (0,\infty),\\
		z(x,t)=0,~~\mathrm{on}~ \Gamma_{D}\times (0,\infty),\\
		\frac{\partial z}{\partial {\bf n}}(x, t)=0,~~\mathrm{on}~ \Gamma_{N}\times (0,\infty), \\
	z(\cdot,0)=z_{0}(\cdot) \in L^{2}(\Omega), \vspace{-0.3cm}
	\end{array}
\end{equation}
where $q\in\mathbb{R}$ is a constant reaction coefficient, 
$\tau_{u}(t)$ is a known input delay which is upper bounded by $\tau_{M,u}$. 
${\bf b}=[b_{1},\dots,b_{d}]^{\mathrm{T}} \in (L^{2}(\Omega))^{d}$, $u(t)=[u_{1}(t),\dots,u_{d}(t)]^{\mathrm{T}}$ is the control input to be designed later.
Let $\delta > 0$.  From \eqref{eigenvalue2}, it follows that there exists $N_{0}\in \mathbb{N}$ such that \vspace{-0.3cm}
 \begin{equation}\label{eq4}
	-\lambda_{n}+q+\delta < 0, ~n>N_{0}, \vspace{-0.3cm}
\end{equation}
where $N_0$ is the number of modes used for the controller design. Let $N \in  \mathbb{N}$, $N \geq N_0$, where $N$ will be the dimension of the observer.
Let $d$ be the maximum of the geometric multiplicities of $\lambda_{n}$, $n=1,\dots, N_{0}$. 
Assume the following delayed non-local measurement: \vspace{-0.32cm}
\begin{equation}\label{measure133} 
	\begin{array}{ll}
		y(t)=\langle {\bf c}, z(\cdot,t-\tau_{y}(t)) \rangle , ~~ t-\tau_{y}(t) \geq 0,\\
		y(t)=0, t-\tau_{y}(t)<0,~{\bf c}=[c_1,\dots, c_d]^{\mathrm{T}} \in (L^{2}(\Omega))^{d}, ~~~\vspace{-0.14cm}
	\end{array} 
\end{equation}
where  $\tau_{y}(t)$ is a known measurement delay which is upper bounded by $\tau_{M,y}$. The controller construction will follow \cite{katz2020constructive} for 1D case (where only simple eigenvalues appear), but the single-input and single-output as in \cite{katz2020constructive} are not applicable to the 2D case due to the existence of multiple eigenvalues (the system is uncontrollable and unobservable).  
Here we introduce multi-input $u(t)$ and multi-output \eqref{measure133} with ${\bf b}$, ${\bf c}$ satisfying Assumption \ref{assump1} (see below) to manage with the controllability and observability.

\vspace{-0.14cm}
We treat two classes of input/output delays: continuously differentiable delays and sawtooth delays that correspond to network-based control. For the case of continuously differentiable delays, we assume that $\tau_{u}(t)$ and $\tau_{y}(t)$ are lower bounded by $\tau_{m}>0$. 
This assumption is employed for well-posedness only. 
As in \cite{katz2021delayed,liu2014delay}, 
we assume that there exists a unique $t_{*}\in [\tau_{m}, \min\{\tau_{M,y},\tau_{M,u}\}]$ such that $t-\tau(t)<0$ if $t<t_{*}$ and $t-\tau(t)\geq 0$ if $t\geq t_{*}$ for $\tau(t)\in\{\tau_{u}(t),\tau_{y}(t)\}$.
For the case of sawtooth delays, $\tau_{y}(t)$ and $\tau_{u}(t)$ are induced by two networks: from sensor to controller and from controller to actuator, respectively (see Section 7.5 in \cite{fridman2014introduction}). 
Henceforth the dependence of $\tau_{y}(t)$ and $\tau_{u}(t)$ on $t$ will be suppressed to shorten notations.

\vspace{-0.14cm}

We present the solution to \eqref{pde1aab} as \vspace{-0.3cm}
\begin{equation}\label{eq2a00}
\setlength{\arraycolsep}{0.1pt} \begin{array}{ll}
	z(x,t)\overset{L^{2}}{=}\sum_{n=1}^{\infty}z_{n}(t)\phi_{n}(x),~
	z_{n}(t)=\langle z(\cdot,t), \phi_{n}\rangle, \vspace{-0.3cm}
\end{array}
\end{equation}
where $\{\phi_{n}\}_{n=1}^{\infty}$ are corresponding eigenfunctions of eigenvalues \eqref{eigenvalue2}.
Differentiating $z_{n}$ in \eqref{eq2a00} and applying Green's first identity,  we obtain \vspace{-0.33cm}
\begin{equation}\label{obeq33aa00}
\begin{array}{ll}
	\dot z_{n}(t) = (-\lambda_{n}+q)z_{n}(t)+{\bf b}_{n}^{\mathrm{T}}u(t-\tau_{u}),~t\geq 0,\\
		z_{n}(0)=\langle z(\cdot,0),\phi_{n}\rangle,
		~~{\bf b}_{n}= \langle {\bf b}, \phi_{n} \rangle \in \mathbb{R}^{d}. \vspace{-0.3cm}
	\end{array}
\end{equation}
We construct a $N$-dimensional observer of the form \vspace{-0.32cm}
\begin{equation}\label{eqaa100}
\begin{array}{ll}
  \hat{z}(x,t)=\sum_{n=1}^{N}\hat{z}_{n}(t)\phi_{n}(x),~~N>N_{0}, \vspace{-0.25cm}
  \end{array}
\end{equation}
where $\hat{z}_{n}(t)$ $(1\leq n\leq N)$ satisfy \vspace{-0.25cm}
\begin{equation}\label{observer22300}
  \begin{array}{ll}
  \dot{\hat{z}}_{n}(t)=(-\lambda_{n}+q)\hat{z}_{n}(t)+{\bf b}_{n}^{\mathrm{T}}u(t-\tau_u)\\
  ~~~~~~~~~~~-l_{n} \left[ \langle {\bf c}, \hat{z}(\cdot,t-\tau_{y}) \rangle  -y(t)\right]  , ~t>0, \\
  \hat{z}_{n}(0)=0,~~t\leq 0,  \vspace{-0.3cm}
  \end{array}
\end{equation}
with $y(t)$ in \eqref{measure133}, observer gains $l_{n}\in \mathbb{R}^{1 \times d}$, $1\leq n \leq N_0$ being designed later and $l_{n}=0_{1\times d}$ for $N_{0}<n\leq N$.

Introduce the notations \vspace{-0.3cm}
\begin{equation}\label{symble11000}
\begin{array}{ll}
	A_{0}= \mathrm{diag}\{-\lambda_{n} +q\}_{n=1}^{N_{0}},  
	A_{1}= \mathrm{diag}\{-\lambda_{n} +q\}_{n=N_{0}+1}^{N},\\
 {\bf c}_{n}= \langle {\bf c},\phi_{n} \rangle,
{\bf C}_{0}=[{\bf c}_{1}, \dots, {\bf c}_{N_{0}}],  {\bf C}_{1}=[{\bf c}_{N_{0}+1}, \dots, {\bf c}_{N}],\\
{\bf B}_{0}=[ {\bf b}_{1}, \dots, {\bf b}_{N_{0}} ]^{\mathrm{T}},~~	 {\bf B}_{1}= [ {\bf b}_{N_{0}+1}, \dots, {\bf b}_{N} ]^{\mathrm{T}}.  \vspace{-0.3cm}
\end{array}	
\end{equation}
We rewrite $A_{0}$ as: \vspace{-0.3cm}
\begin{equation}\label{A0A0A0}
\begin{array}{ll}
	A_{0}= \mathrm{diag}\{\tilde{A}_{1}, \dots, \tilde{A}_{p}\}, \\
	\tilde{A}_{j}=\mathrm{diag}\{-\lambda_{j}+q, \dots, -\lambda_{j}+q\}\in\mathbb{R}^{n_{j}\times n_{j}},\\
	\lambda_{k} \neq \lambda_{j} ~~\mathrm{iff}~~k\neq j, ~~ k,j=1,\dots,p,   \vspace{-0.3cm}
\end{array}	
\end{equation}
 where 
 $n_{1},\dots, n_{p}$ are positive integers such that $n_{1}+\dots+n_{p}=N_{0}$. Clearly, $n_{j}\leq d$, $j=1,\dots,p$ and there exists at least one $\jmath\in \{1,\dots,p\}$ such that $n_{\jmath}= d$. 
According to the partition of \eqref{A0A0A0}, we rewrite ${\bf B}_{0}$ and ${\bf C}_{0}$  as \vspace{-0.3cm}
\begin{equation*}
\begin{array}{ll}
	{\bf B}_{0}=\mathrm{col}\{ B_{j}\}_{j=1}^{p}, ~~B_{j}\in\mathbb{R}^{n_{j}\times d}, \\ 
	{\bf C}_{0}=[C_{1}, \dots, C_{p}], ~~C_{j}\in\mathbb{R}^{d\times n_{j}}. \vspace{-0.3cm}
\end{array}	
\end{equation*} \vspace{-0.6cm}
\begin{assumption}\label{assump1}
Let $\mathrm{rank}(B_{j}) =n_{j}$ and $\mathrm{rank}(C_{j})= n_{j}$,  $j=1,\dots,p$.
\end{assumption}\vspace{-0.25cm}
\begin{lemma}\label{proposition2.1}
	Under Assumption \ref{assump1},  the pair $(A_{0},{\bf B}_{0})$ is controllable and the pair $(A_{0},{\bf C}_{0})$ is observable.
\end{lemma}
\vspace{-0.7cm}
\begin{pf}
	The proof is inspired by Lemma 7.2 of \cite{meng2022boundary}. Assume that the pair $(A_{0},{\bf C}_{0})$ is not observable. By the Hautus test (see \cite[Remark 1.5.2]{tucsnak2009observation}), there exist $0\neq \nu\in\mathbb{R}^{N_{0}}$ and $j\in\{1,\dots,p\}$ such that \vspace{-0.3cm}
	\begin{equation}\label{2.10aa}
		\begin{array}{ll}
			A_{0}\nu=\lambda_{j}\nu,~~~~{\bf C}_{0} \nu=0. \vspace{-0.3cm}
	\end{array}
	\end{equation}
	Without loss of generality, we suppose that $\nu=\mathrm{col}\{\nu_{1},\dots, \nu_{p}\}$, where $\nu_{j}=[\nu^{(1)}_{j}, \dots, \nu^{(n_{j})}_{j} ]^{\mathrm{T}}$. Then \eqref{2.10aa} becomes $A_{0}\nu - \lambda_{j}\nu= \mathrm{col}\{(\lambda_{k}-\lambda_{j})\nu_{k}\}_{k=1}^{p}=0$ and $\sum_{k=1}^{p}C_{k}\nu_{k}=0$, 
	which implies $\nu_{k}=0$ for $k\neq j$ and $ C_{j}\nu_{j}=0$. Since $\mathrm{rank}(C_{j})= n_{j}$, we have $\nu_{j}=0$. This contradicts to the fact $\nu\neq 0$. Therefore, pair $(A_{0},{\bf C}_{0})$ is observable. The controllability of $(A_{0},{\bf B}_{0})$ follows similarly. 
\end{pf} 

\vspace{-0.4cm}
Under Assumption \ref{assump1}, we can let  $L_{0}= \mathrm{col}\{l_{1},\dots , l_{N_{0}}\}\in \mathbb{R}^{N_0 \times d}$ and $K_{0}\in \mathbb{R}^{d\times N_{0}}$ satisfy  \vspace{-0.32cm} 
\begin{subequations}\label{bL000}
\begin{align}
&P_{o}(A_{0}-L_{0}{\bf C}_{0})+(A_{0}-L_{0}{\bf C}_{0})^{\mathrm{T}}P_{o}<-2\delta P_{o},   \label{bL00011}\\
& P_{c}(A_{0}- {\bf B}_{0}K_{0})+(A_{0}- {\bf B}_{0}K_{0})^{\mathrm{T}}P_{c}   \leq -2\delta P_{c},  \label{bL00022} 
 \end{align} 
\end{subequations}
for $0<P_{o},P_{c}\in \mathbb{R}^{N_{0}\times N_{0}}$. We propose a controller of the form \vspace{-0.45cm}
\begin{equation}\label{eq600}
\setlength{\arraycolsep}{0.1pt}\begin{array}{ll}
	u(t)=-K_{0}\hat{z}^{N_{0}}(t), ~~ \hat{z}^{N_{0}}=[\hat{z}_{1},\dots, \hat{z}_{N_{0}}]^{\mathrm{T}}. \vspace{-0.3cm}
\end{array}
\end{equation}

\vspace{-0.1cm} 
For well-posedness of closed-loop system \eqref{pde1aab}, \eqref{observer22300} with control input \eqref{eq600}, we consider the state $\xi(t)= \mathrm{col}\{z(\cdot,t), \hat{z}^{N}(t)\}$, where $\hat{z}^{N}(t)= \mathrm{col}\{\hat{z}_{n}(t)\}_{n=1}^{N}$. The closed-loop system can be presented as \vspace{-0.35cm}
\begin{equation}\label{abstractsystem}	
\setlength{\arraycolsep}{0.8pt} {\scriptsize \begin{array}{ll}
		\frac{\mathrm{d}}{\mathrm{d}t}\xi(t)+ \mathrm{diag}\{\mathcal{A},  \mathcal{A}_{0}\}\xi(t)
		={ \tiny \left[\begin{array}{cc}
		qz(\cdot,t)+f_{1}(t-\tau_{u})\\
	f_{2}(t-\tau_{u}) + f_{3}(t-\tau_{y})\end{array} \right]},\\
	\mathcal{A}_{0}= \mathrm{diag}\{-A_{0}, -A_{1}\},~f_{1}(t)= - {\bf b}^{\mathrm{T}}(\cdot)K_{0}\hat{z}^{N_{0}}(t),\\
	f_{2}(t)= - {\bf B}K_{0}\hat{z}^{N_{0}}(t)	,~{\bf B}=\mathrm{col}\{{\bf B}_{1},{\bf B}_{2} \}, ~ {\bf C}=[{\bf C}_{0},{\bf C}_{1}],\\
	f_{3}(t)=-{ \tiny \left[\begin{array}{cc}
		L_{0}\\
		0_{(N-N_{0})\times d}\end{array} \right]} [ {\bf C}\hat{z}^{N}(t) - \langle {\bf c}, z(\cdot,t) \rangle  ], \vspace{-0.32cm}
	\end{array} }
\end{equation}
where $\mathcal{A}$ is defined in \eqref{operatorA}. 
We begin with continuously differentiable delays. By using Theorems 6.1.2 and 6.1.5 in \cite{pazy2012semigroups}  together with the step method on intervals $[0,t_{*}]$, $[t_{*},(s+1)\tau_{m}]$, $[(s+1)\tau_{m},(s+2)\tau_{m}]$, $\dots$, where $s\in\mathbb{N}$ satisfies $s\tau_{m}\leq t_{*} < (s+1)\tau_{m}$
(see arguments similar to the well-posedness in Section 3 of \cite{katz2021delayed}), 
we obtain that for any initial value $\xi(0)=[z_{0}(\cdot), 0]^{\mathrm{T}}\in \mathcal{D}(\mathcal{A})\times \mathbb{R}^{N}$, the closed-loop system \eqref{abstractsystem} has a unique classical solution \vspace{-0.35cm}
\begin{equation}\label{regularity}
	{\scriptsize \begin{array}{ll}
		\xi \in C([0,\infty), L^{2}(\Omega)\times \mathbb{R}^{N})\cap C^{1}([0,\infty)\backslash J, L^{2}(\Omega)\times \mathbb{R}^{N}),\\
		\xi(t)\in \mathcal{D}(\mathcal{A}) \times \mathbb{R}^{N}, ~~\forall t\geq 0, \vspace{-0.35cm}
	\end{array}}
\end{equation}
where $J=\{t_{*},(s+1)\tau_{m},(s+2)\tau_{m},\dots\}$. 
The well-posedness for sawtooth delays follows similarly.

\vspace{-0.2cm}
\subsection{Stability analysis and main results}\vspace{-0.3cm}
Let $e_{n}(t)=z_{n}(t)-\hat{z}_{n}(t)$, $1\leq n \leq N$ 
be the estimation error. The last term on the right-hand side of \eqref{observer22300} can be presented as \vspace{-0.3cm}
\begin{equation}\label{tail100aa}
	\begin{array}{ll}
		\sum_{n=1}^{N}{\bf c}_{n} \hat{z}_{n}(t-\tau_{y})  -y(t) \\
		=-\sum_{n=1}^{N}{\bf c}_{n}e_{n}(t-\tau_{y}) -\zeta(t-\tau_{y}),\\
		\zeta(t)=\sum_{n=N+1}^{\infty}{\bf c}_{n}z_{n}(t). \vspace{-0.3cm}
	\end{array}
\end{equation}
From \eqref{obeq33aa00}, \eqref{observer22300}, \eqref{tail100aa},   the error system has the form \vspace{-0.3cm}
\begin{equation}\label{error00}
  \begin{array}{ll}
  \dot{e}_{n}(t)=(-\lambda_{n}+q)e_{n}(t)   -l_{n}  \sum_{i=1}^{N}{\bf c}_{i}e_{i}(t-\tau_{y})\\
  ~~~~~~~~~~~-l_{n} \zeta(t-\tau_{y}) , ~1\leq n \leq N. \vspace{-0.25cm}
  \end{array}
\end{equation}
Denote \vspace{-0.3cm}
\begin{equation}\label{symbol00}
	{\scriptsize \begin{array}{ll}
		  \hat{z}^{N-N_{0}}(t)= [\hat{z}_{N_{0}+1}(t), \dots,\hat{z}_{N}(t)]^{\mathrm{T}}, \\
   e^{N_{0}}(t) = [e_{1}(t), \dots,e_{N_{0}}(t)]^{\mathrm{T}},\\
   e^{N-N_{0}}(t) = [e_{N_{0}+1}(t), \dots,e_{N}(t)]^{\mathrm{T}}, \\
    X_{0}(t)=\mathrm{col}\{\hat{z}^{N_{0}}(t),e^{N_{0}}(t)\}, ~~~~~\mathcal{K}_{0}=  [ K_{0}, 0_{d\times N_{0}} ], \\
    {\tiny F_{0}=\left[
               \begin{array}{cccc}
                 A_{0}-{\bf B}_{0}K_{0} & L_{0}{\bf C}_{0} \\
                 0     & A_{0}-L_{0}{\bf C}_{0} 
               \end{array}
             \right]}, ~ \mathcal{L}_{0}=\mathrm{col}\{L_{0}, -L_{0} \}, \\
 \nu_{\tau_{u}}(t)= \hat{z}^{N_0}(t)-\hat{z}^{N_0}(t-\tau_{u}),  ~~\mathcal{B}_{0}=  \mathrm{col}\{ {\bf B}_{0}, 0_{N_{0}\times d } \},\\
   \nu_{\tau_{y}}(t)= X_0(t)-X_0(t-\tau_{y}),~~~~~\mathcal{C}_{0}=[{\bf C}_{0},0_{d\times N_{0}}]. \vspace{-0.3cm}
	\end{array} }
\end{equation}
From \eqref{error00}, we have  
$e^{N-N_{0}}(t)=\mathrm{e}^{A_{1}t}e^{N-N_{0}}(0)$. 
By \eqref{observer22300}, \eqref{eq600}, \eqref{error00} and substituting $e^{N-N_{0}}(t-\tau_{y})=\mathrm{e}^{-A_{1}\tau_{y}}e^{N-N_{0}}(t)$, we obtain the reduced-order closed-loop system \vspace{-0.3cm}
\begin{subequations}\label{eqXX00}
{\scriptsize \begin{align}
&\dot{X}_{0}(t)= F_{0}X_{0}(t) + \mathcal{B}_{0}K_{0}\nu_{\tau_{u}}(t)- \mathcal{L}_{0}\mathcal{C}_{0}\nu_{\tau_{y}}(t)  \label{eqXX0011}\\
& ~~~~~~~~~+ \mathcal{L}_{0} \zeta(t-\tau_{y})  
 +\mathcal{L}_{0}\mathbf{C}_{1}\mathrm{e}^{-A_{1}\tau_{y}}e^{N-N_{0}}(t),  \nonumber \\
& \dot z_{n}(t)=(-\lambda_{n}+q)z_{n}(t) - {\bf b}^{\mathrm{T}}_{n}\mathcal{K}_{0}X_{0}(t-\tau_{y}),~n>N,  \label{eqXX0022} 
\end{align}}
\end{subequations}
where $\zeta(t)$ is defined in \eqref{tail100aa}. Note that $\zeta(t)$ does not depend on $\hat{z}^{N-N_{0}}(t)$ which satisfies \vspace{-0.3cm}
\begin{equation}\label{hatzNN0}
	\begin{array}{ll}
		\dot{\hat{z}}^{N-N_{0}}(t)=A_{1}\hat{z}^{N-N_{0}}(t)-{\bf B}_{1}\mathcal{K}_{0}X_{0}(t-\tau_u), 
	\end{array} \vspace{-0.3cm}
\end{equation}
and is exponentially decaying (since $A_1$ defined in \eqref{symble11000} is stable due to \eqref{eq4}) provided $X_{0}(t)$ is exponentially decaying. Therefore, for stability of \eqref{pde1aab} under the control law \eqref{eq600}, it is sufficient to show the stability of the reduced-order system \eqref{eqXX00}. 
The latter can be considered as a singularly perturbed system with the slow sate $X_0 (t)$ and the fast infinite-dimensional state $z_n(t)$, $n>N$.

\vspace{-0.2cm}

For exponential $L^{2}$-stability of the closed-loop system \eqref{eqXX00}, we consider the following vector Lyapunov functional \vspace{-0.3cm}
\begin{equation}\label{FSV}
{\scriptsize \begin{array}{ll}
	V(t)= [V_{0}(t) , ~~V_{\mathrm{tail}}(t)]^{\mathrm{T}},
	V_{\mathrm{tail}}(t)=\sum_{n=N+1}^{\infty}z^{2}_{n}(t), \\
	V_{0}(t)=V_{P}(t) + V_{y}(t)+V_{u}(t) + V_{e}(t), \\
	V_{P}(t)=|X_{0}(t)|_{P}^{2},~~~V_{e}(t)=  p_{e}|e^{N-N_{0}}(t)|^2, \\
	V_{y}(t)= \int_{t-\tau_{M,y}}^{t}\mathrm{e}^{2\delta(s-t)}|X_{0}(s)|^{2}_{S_{y}}\mathrm{d}s, \\
	~~~~~~~+\tau_{M,y} \int_{-\tau_{M,y}}^{0}\int_{t+\theta}^{t}\mathrm{e}^{2\delta(s-t)}|\dot{X}_{0}(s)|^{2}_{R_{y}}\mathrm{d}s\mathrm{d}\theta,\\
	V_{u}(t)= \int_{t-\tau_{M,u}}^{t}\mathrm{e}^{2\delta(s-t)}|\mathcal{K}_{0}X_{0}(s)|^{2}_{S_{u}}\mathrm{d}s \\
	~~~~~~~+\tau_{M,u} \int_{-\tau_{M,u}}^{0}\int_{t+\theta}^{t}\mathrm{e}^{2\delta(s-t)}|\mathcal{K}_{0}\dot{X}_{0}(s)|^{2}_{R_{u}}\mathrm{d}s\mathrm{d}\theta, \vspace{-0.2cm}
\end{array} }
\end{equation}
where
$0<P, S_{y}, R_{y} \in \mathbb{R}^{2N_{0}\times 2N_{0}}$ and $0<S_{u},R_{u}\in \mathbb{R}^{d\times d}$. Here $V_{y}(t)$ is used to compensate $\nu_{\tau_{y}}(t)$, $V_{u}(t)$ is used to compensate $\nu_{\tau_{u}}(t)$, and $V_{e}(t)$ is used to compensate $e^{N-N_{0}}(t)$. 
To compensate $\zeta(t-\tau_{y})$ we will use vector Halanay's inequality and
the following Cauchy-Schwarz inequality: \vspace{-0.35cm}
\begin{equation}\label{CSineqaa}
\begin{array}{ll}
	|\zeta(t)|^{2}\leq \|{\bf c}\|^{2}_{N} \sum_{n=N+1}^{\infty}z^{2}_{n}(t),\\
	\|{\bf c}\|^{2}_{N}:=\sum_{j=1}^{d}\|c_{j}\|^{2}_{N}=\sum_{n=N+1}^{\infty}|{\bf c}_{n}|^{2}.\vspace{-0.3cm}
\end{array}	
\end{equation}
As explained in Remark \ref{remkdelta1} below, compared to the classical Halanay's inequality, 
the vector one allows to use smaller $\delta$ in $V_y$ and $V_u$ in the stability analysis essentially improving results in the numerical examples for comparatively large $N$.


Differentiation of $V_{\mathrm{tail}}(t)$ along \eqref{eqXX0022} gives \vspace{-0.3cm}
\begin{equation}\label{Vtail}
	\begin{array}{ll}
		\dot{V}_{\mathrm{tail}}(t)= \sum_{n=N+1}^{\infty}2(-\lambda_{n}+q)z^{2}_{n}(t) \\
		~~~~~~~~~~~~~-\sum_{n=N+1}^{\infty}2z_{n}(t){\bf b}^{\mathrm{T}}_{n}\mathcal{K}_{0}X(t-\tau_{u}). \vspace{-0.2cm}
	\end{array}
\end{equation}
Let $\alpha >0$. Applying Young's inequality we arrive at \vspace{-0.35cm}
\begin{equation}\label{byoung}
\setlength{\arraycolsep}{1.3pt}	\begin{array}{ll}
		-\sum_{n=N+1}^{\infty}2z_{n}(t){\bf b}^{\mathrm{T}}_{n}\mathcal{K}_{0}X(t-\tau_{u}) \\
		\leq \frac{\|{\bf b}\|^{2}_{N} }{\alpha}  X^{\mathrm{T}}(t-\tau_{u})\mathcal{K}^{\mathrm{T}}_{0} \mathcal{K}_{0}X(t-\tau_{u}) \\
		~~+ \alpha \sum_{n=N+1}^{\infty} z^{2}_{n}(t),~~
		\|{\bf b}\|^{2}_{N}: = \sum_{i=1}^{d}\|b_{i}\|_{N}^{2}. \vspace{-0.25cm}
	\end{array}
\end{equation}
From \eqref{Vtail} and \eqref{byoung}, we have \vspace{-0.3cm}
\begin{equation}\label{VXVXV}
	 \begin{array}{ll}
		\dot{V}_{\mathrm{tail}}(t) + [2\lambda_{N+1}-2q-\alpha ]V_{\mathrm{tail}}(t) \\
		\leq   \frac{\|{\bf b}\|^{2}_{N} }{\alpha} |\mathcal{K}_{0}X(t-\tau_{u}) |^2
		\leq \beta V_{0}(t-\tau_{u}) \vspace{-0.2cm}
	\end{array}
\end{equation}
provided \vspace{-0.2cm}
\begin{equation}\label{LMI11}
	\begin{array}{ll}
		\frac{\|{\bf b}\|^{2}_{N} }{\alpha} \mathcal{K}^{\mathrm{T}}_{0}\mathcal{K}_{0}  < \beta P. 
	\end{array}
\end{equation}
Let $\beta_{0}=\alpha\beta$.
By Schur complement, we find that \eqref{LMI11} holds iff 
\begin{equation}\label{LMI11abc}
	\begin{array}{ll}
		{\scriptsize \left[
               \begin{array}{ccc}
               -P  &  \mathcal{K}_{0}^{\mathrm{T}} \\
               *   & - {\beta_{0} \over \|{\bf b}\|^{2}_{N}}I
               \end{array}
           \right]}<0. \vspace{-0.01cm}
	\end{array}
\end{equation}
 Let \vspace{-0.3cm}
\begin{equation*}
\setlength{\arraycolsep}{0.5pt} {\scriptsize \begin{array}{ll}
	\varepsilon_{y}=e^{-2\delta\tau_{M,y}}, ~
	\theta_{\tau_y}(t)=e^{N_{0}}(t-\tau_{y})-e^{N_{0}}(t-\tau_{M,y}), \\
	\varepsilon_{u}=e^{-2\delta\tau_{M,u}},~\theta_{\tau_u}(t)=\hat{z}^{N_{0}}(t-\tau_{u})-\hat{z}^{N_{0}}(t-\tau_{M,u}). \vspace{-0.3cm}
\end{array}	}
\end{equation*}
Differentiation of $V_{0}(t)$ along \eqref{eqXX0011} gives \vspace{-0.3cm}
\begin{equation}\label{dotVX}
{\scriptsize	\begin{array}{ll}
		\dot{V}_{0}(t)+ 2\delta V_{0}(t)\leq X_{0}^{\mathrm{T}}(t)[PF_{0}+F_{0}^{\mathrm{T}}P+2\delta P]X_{0}(t)\\
		+2X_{0}^{\mathrm{T}}(t)P[ \mathcal{B}_{0}K_{0}\nu_{\tau_{u}}(t)- \mathcal{L}_{0}\mathcal{C}_{0}\nu_{\tau_{y}}(t)  + \mathcal{L}_{0} \zeta(t-\tau_{y})]\\
		+2X_{0}^{\mathrm{T}}(t)P\mathcal{L}_{0}\mathbf{C}_{1}\mathrm{e}^{-A_{1}\tau_{y}}e^{N-N_{0}}(t) \\
		+|X_{0}(t)|^{2}_{S_{y}}  - \varepsilon_{y} |X_{0}(t)-\nu_{\tau_{y}}(t)-\theta_{\tau_{y}}(t) |^{2}_{S_{y}}\\
		+\tau^{2}_{M,y}|\dot{X}_{0}(t)|^{2}_{R_{y}} -\varepsilon_{y}\tau_{M,y}\int_{t-\tau_{M,y}}^{t}|\dot{X}_{0}(s)|^{2}_{R_{y}}\mathrm{d}s \\
		+|\mathcal{K}_{0} X_{0}(t)|^{2}_{S_{u}} -\varepsilon_{u} |\mathcal{K}_{0}X_{0}(t)-K_{0}\nu_{\tau_u}(t)-K_{0}\theta_{\tau_u}(t) |^{2}_{S_{u}}\\
		+\tau^{2}_{M,u}|\mathcal{K}_{0} \dot{X}_{0}(t)|^{2}_{R_{u}} -\varepsilon_{u}\tau_{M,u}\int_{t-\tau_{M,u}}^{t}|\mathcal{K}_{0}\dot{X}_{0}(s)|^{2}_{R_{u}}\mathrm{d}s\\
		+ 2p_{e}(e^{N-N_{0}}(t))^{\mathrm{T}}[A_{1}+\delta I] e^{N-N_{0}}(t). \vspace{-0.3cm}
	\end{array} }
\end{equation}
Let $G_{y}\in \mathbb{R}^{2N_0\times 2N_0}$ and $G_{u}\in \mathbb{R}^{d\times d}$  satisfy \vspace{-0.3cm}
\begin{equation}\label{G1G2}
{  \tiny \left[
               \begin{array}{cc}
               R_{y} & G_{y}  \\
                 * &  R_{y}
               \end{array}
           \right]\geq 0, ~	\left[
               \begin{array}{cc}
               R_{u} & G_{u}  \\
                 * &  R_{u}
               \end{array}
           \right]} \geq 0. \vspace{-0.3cm}
\end{equation}
Applying Jensen's and Park's inequalities (see, e.g., \cite[Section 3.6.3]{fridman2014introduction}), we obtain  for 
$\xi_{y}(t)=\mathrm{col}\{ \nu_{\tau_y}(t), \theta_{\tau_y}(t)\}$, $\xi_{u}(t)=\mathrm{col}\{ K_{0} \nu_{\tau_u}(t), K_{0}\theta_{\tau_u}(t)\}$, \vspace{-0.3cm}
\begin{equation}\label{FFF}
\setlength{\arraycolsep}{0.3pt} {\scriptsize \begin{array}{ll}
- \tau_{M,y}\int_{t-\tau_{M,y}}^{t}|\dot{X}_{0}(s)|^{2}_{R_{y}}\mathrm{d}s   \leq

        -\xi^{\mathrm{T}} _{y}(t) { \tiny \left[
               \begin{array}{cc}
               R_{y} & G_{y}  \\
                 * &  R_{y}
               \end{array}
           \right]}
\xi_{y}(t), \\
 - \tau_{M,u}\int_{t-\tau_{M,u}}^{t}|\mathcal{K}_{0}\dot{X}_{0}(s)|^{2}_{R_{u}}\mathrm{d}s  \leq
    -  \xi^{\mathrm{T}}_{u}(t)  {\tiny \left[
               \begin{array}{cc}
               R_{u} & G_{u}  \\
                 * &  R_{u}
               \end{array}
           \right]}  \xi_{u}(t).  \vspace{-0.3cm}
 \end{array}	}
\end{equation}
Let $\eta(t)=\mathrm{col}\{ X_{0}(t), \zeta(t-\tau_{y}), \xi_{y}(t),  \xi_{u}(t),e^{N-N_{0}}(t)\}$.
Substituting \eqref{FFF} into \eqref{dotVX},  we get for $\delta_{1}>0$,  \vspace{-0.3cm}
\begin{equation}\label{bLVLV}
 \begin{array}{ll}
	\dot{V}_{0}(t)+ 2\delta V_{0}(t) - 2\delta_{1}V_{\mathrm{tail}}(t-\tau_{y})
	\\
\overset{\eqref{CSineqaa}}	\leq \dot{V}_{X}(t)+ 2\delta V_{X}(t) - \frac{2\delta_{1}}{ \|{\bf c}\|^{2}_{N} }|\zeta(t-\tau_{y})|^{2}  \\
~	\leq \eta^{\mathrm{T}}(t)\Phi\eta(t)  \leq 0 \vspace{-0.3cm}
\end{array}	
\end{equation}
provided \vspace{-0.3cm}
\begin{equation}\label{bLMI1}
	\begin{array}{ll}
\Phi= {  \small \left[
               \begin{array}{c|c}
               \Phi_{0} & P\mathcal{L}_{0}\mathbf{C}_{1}\mathrm{e}^{-A_{1}\tau_{y}}   \\ \hline
                 * &  2p_{e}(A_1 +\delta I)
               \end{array}
           \right]} \\
    ~~~~~~ + \Lambda^{\mathrm{T}}[\tau_{M,y}^{2}R_{y}  +  \tau_{M,u}^{2}\mathcal{K}^{\mathrm{T}}_{0}R_{u}\mathcal{K}_{0}]\Lambda \leq 0,  \vspace{-0.3cm}
\end{array} 
\end{equation}
where \vspace{-0.3cm}
\begin{equation*}
	\setlength{\arraycolsep}{2pt}{\scriptsize \begin{array}{ll}
	\Phi_{0}={\tiny \left[
               \begin{array}{c|c|c}
             \begin{array}{cc}
                \Omega_{0}  & P\mathcal{L}_{0}  \\
                * &  -\frac{2\delta_{1}}{\|{\bf c}\|^{2}_{N}  } I
               \end{array}   & \begin{array}{cc}
              \Omega_1   & \varepsilon_{y}S_{y}  \\
                0 &  0
               \end{array}
              &   \begin{array}{cc}
               \Omega_2   & \varepsilon_{u}\mathcal{K}_{0}^{\mathrm{T}} S_{u} \\
                0 &  0
               \end{array}  \\ \hline
                 * & \Omega_{y}  & 0 \\\hline
                 * &*&  \Omega_{u}

               \end{array}
           \right]}, \\
	    \Omega_{0}= PF_{0}+F_{0}^{\mathrm{T}}P+2\delta P + (1-\varepsilon_{y})S_{y}+(1-\varepsilon_{u})\mathcal{K}^{\mathrm{T}}_{0}S_{u}\mathcal{K}_{0},  \\
	      \Omega_{1}=\varepsilon_{y}S_{y}-P\mathcal{L}_{0}\mathcal{C}_{0},~~~\Omega_{2}=P\mathcal{B}_{0} +\varepsilon_{u}\mathcal{K}_{0}^{\mathrm{T}}S_{u},
	\end{array} }
\end{equation*}\vspace{-0.9cm}
\begin{equation}\label{bLMI1a}
	\setlength{\arraycolsep}{2pt}{\scriptsize \begin{array}{ll}
	  \Lambda= [\Lambda_{0}, ~ \mathcal{L}_{0}\mathbf{C}_{1}\mathrm{e}^{-A_{1}\tau_{y}}], ~\Lambda_{0}= [F_{0}, \mathcal{L}_{0}, -\mathcal{L}_{0}\mathcal{C}_{0}, 0, \mathcal{B}_{0}, 0 ],\\	
     \Omega_{J} = { \left[
               \begin{array}{cc}
               -\varepsilon_{J}(S_{J}+R_{J}) & ~-\varepsilon_{J}(S_{J}+G_{J})  \\
                 * &  ~-\varepsilon_{J}(S_{J}+R_{J})
               \end{array}
           \right]}, J\in\{y,u\}. \vspace{-0.3cm}
	\end{array} }
\end{equation}
We now show the feasibility of \eqref{bLMI1} for large $N$. Since $A_{1}+\delta I<0$ due to \eqref{eq4}, by Schur complement for $p_{e}\rightarrow \infty$, we obtain that the feasibility of \eqref{bLMI1} holds iff \vspace{-0.3cm}
\begin{equation}\label{bLMI1ab}
	\begin{array}{ll}
\Phi_{0}+ \Lambda_{0}^{\mathrm{T}}[\tau_{M,y}^{2}R_{y}  +  \tau_{M,u}^{2}\mathcal{K}^{\mathrm{T}}_{0}R_{u}\mathcal{K}_{0}]\Lambda_{0} \leq 0. \vspace{-0.3cm}
	\end{array}
\end{equation}
From \eqref{VXVXV} and \eqref{bLVLV}, we have \vspace{-0.3cm}
\begin{equation}\label{dotV}
	\begin{array}{ll}
		\dot{V}(t)
        \leq   {\scriptsize \left[
               \begin{array}{cc}
               -2\delta &0 \\
               0 & -2\lambda_{N+1}+2q+\frac{1}{\alpha}
               \end{array}
           \right] }V(t) \\
           ~~~~~~~+  {\scriptsize \left[
               \begin{array}{cc}
               0 &2\delta_{1} \\
               0 & 0
               \end{array}
           \right] }V(t-\tau_{y}) +  {\scriptsize \left[
               \begin{array}{cc}
               0 &0 \\
               \beta & 0
               \end{array}
           \right] }V(t-\tau_{u}). \vspace{-0.3cm}
	\end{array}
\end{equation}
By vector Halanay's inequality (see Lemma \ref{lemma2}) we have  \vspace{-0.3cm}
\begin{equation}\label{Vstable11}
	\begin{array}{ll}
		|V(t)|\leq D \mathrm{e}^{-2\delta_{0}t}, ~~t\geq 0 \vspace{-0.3cm}
	\end{array}
\end{equation}
for some $\delta_{0}>0$ and $D>0$, provided \vspace{-0.3cm}
\begin{equation}\label{vectorHalanay}
	\begin{array}{ll}
		{ \footnotesize \left[
               \begin{array}{ccc}
              - 2\delta &  2\delta_{1} \\
                \beta    &  - 2\lambda_{N+1}+2q+ \alpha
               \end{array}
           \right]}
	\end{array} \mathrm{is}~~ \mathrm{Hurwitz}. \vspace{-0.3cm}
\end{equation}
By Parseval's equality, we obtain from \eqref{Vstable11} that \vspace{-0.25cm}
\begin{equation}\label{Vstable222}
	 \|z(\cdot,t)\|^{2}_{L^{2}}+\|z(\cdot,t)-\hat{z}(\cdot,t) \|^{2}_{L^{2}}\leq \tilde{D}\mathrm{e}^{-\delta_{0}t},~t\geq 0 \vspace{-0.3cm}
\end{equation}
for some $\tilde{D}>0$. 
Recalling that $\beta_{0}=\alpha\beta$, we find that \eqref{vectorHalanay} holds iff \vspace{-0.35cm}
\begin{equation}\label{Hurwitz2}
	\begin{array}{ll}
		-2(\lambda_{N+1}-q+\delta) + \alpha <0,\\
	{ \footnotesize \left[
               \begin{array}{ccc}
              -2\alpha(\lambda_{N+1}-q) +\frac{\delta_{1}}{\delta}\beta_{0} &  \alpha \\
               *    &  - 1
               \end{array}
           \right]} <0. \vspace{-0.25cm}
	\end{array}
\end{equation}

\vspace{-0.15cm}
For asymptotic feasibility of LMIs \eqref{LMI11abc}, \eqref{G1G2},  \eqref{bLMI1ab}, and \eqref{Hurwitz2} with large $N$ and small $\tau_{M,y}, \tau_{M,u}>0$, let $S_{i}=0$, $G_{i}=0$ for $i\in\{y,u\}$. 
Taking $\tau_{M,y},\tau_{M,u}\rightarrow 0^{+}$, it is sufficient to show \eqref{LMI11abc}, \eqref{Hurwitz2} and \vspace{-0.3cm}
\begin{equation}\label{bLMI1addd}
	\begin{array}{ll}
		{\tiny \left[
               \begin{array}{cc|cc}
               PF_{0}+F_{0}^{\mathrm{T}}P+2\delta P   & P\mathcal{L}_{0}  & -P\mathcal{L}_{0}\mathcal{C}_{0} & P\mathcal{B}_{0} \\
                * & -\frac{2\delta_{1}}{\|{\bf c}\|^{2}_{N}  }I  &   0 &    0  \\ \hline
             *&* & -R_{y}  & 0 \\
              *&   * &*&  -R_{u}
               \end{array}
           \right]}<0. \vspace{-0.3cm}
	\end{array}
\end{equation}
Take $\alpha=\delta=1$, $\delta_{1}=\beta_{0}=N^{\frac{1}{3}}$, $R_{y}=NI$, $R_{u}=NI$. Let $0<P\in \mathbb{R}^{2N_{0}\times 2N_{0}}$ be the solution to the Lyapunov equation $P(F_0+\delta I)+(F_0+\delta I)^{\mathrm{T}}P=-I$. We have $\|P\|=O(1)$, $N\to \infty$.
Substituting above values into \eqref{LMI11abc}, \eqref{Hurwitz2}, \eqref{bLMI1addd} and using Schur complement and the fact that $\lambda_{N}=O(N)$ (see Lemma \ref{lemma}), $\|\mathcal{L}_{0}\|=O(1)$, $\|\mathcal{B}_{0}\|=O(1)$ for $N\rightarrow \infty$, 
 we obtain the feasibility of \eqref{LMI11abc}, \eqref{Hurwitz2} and \eqref{bLMI1addd} for large enough $N$. 
 Fixing such $N$ and using continuity, we have that \eqref{LMI11abc}, \eqref{G1G2},  \eqref{bLMI1} and \eqref{Hurwitz2} are feasible for small enough $\tau_{M,y}, \tau_{M,u}>0$. Summarizing, we arrive at:\vspace{-0.25cm}
\begin{theorem}\label{thm1}
Consider \eqref{pde1aab} with control law \eqref{eq600} and measurement \eqref{measure133}. 
For $\delta>0$, let $N_{0}\in\mathbb{N}$ satisfy \eqref{eq4} and $N\in\mathbb{N}$ satisfy $N\geq N_{0}$. Let Assumption \ref{assump1} hold and $L_{0}$, $K_{0}$ be obtained from \eqref{bL000}.
Given $\tau_{M,y}, \tau_{M,u}>0$ and $\delta_{1}>0$, 
let there exist $0<P,S_{y}, R_{y}\in \mathbb{R}^{2N_{0} \times 2N_{0}}$, $0<S_{u}, R_{u} \in \mathbb{R}^{d \times d}$, $G_{y}\in \mathbb{R}^{2N_0 \times 2N_0}$, $G_{u}\in \mathbb{R}^{d \times d}$ and scalars $\alpha, \beta_{0}>0$  such that LMIs \eqref{LMI11abc}, \eqref{G1G2}, \eqref{bLMI1ab} with $\Phi_{0}$ and $\Lambda_{0}$ given in \eqref{bLMI1a}, and \eqref{Hurwitz2} hold. 
Then the solution $z(x,t)$ to \eqref{pde1aab} subject to the control law \eqref{observer22300}, \eqref{eq600} and the corresponding observer $\hat{z}(x,t)$ given by \eqref{eqaa100} satisfy \eqref{Vstable222} for some $\tilde{D}>0$ and $\delta_{0}>0$.
Moreover, LMIs  \eqref{LMI11abc}, \eqref{G1G2}, \eqref{bLMI1ab}, and \eqref{Hurwitz2} are always feasible for large enough $N$ and small enough $\tau_{M,y},\tau_{M,u}>0$.
\end{theorem} \vspace{-0.2cm}
\begin{remark}\label{remkdelta1}
	Multiplying decision variables $P$, $S_{i}$, $R_{i}$, $G_{i}$ ($i\in\{y,u\}$) in \eqref{LMI11abc}, \eqref{G1G2}, \eqref{bLMI1ab} by $\delta_{1}$ and changing $\beta_{0}$ in \eqref{LMI11abc} and \eqref{Hurwitz2} to ${\beta_{0}\over \delta_{1}}$, 
	we find that the feasibility of LMIs \eqref{LMI11abc}, \eqref{G1G2}, \eqref{bLMI1ab}, and \eqref{Hurwitz2} is independent of $\delta_{1}>0$. The fact also holds true for Theorems \ref{thm3} and \ref{thm2} below. This is different from the classical Halanay inequality (see Remark \ref{remark3}  below) where $\delta_{1}  \leq  \delta$ should not be small to compensate $\zeta(t-\tau_{y})$. 
 However, compared to the classical Halanay inequality, the vector one needs constraint \eqref{LMI11} (i.e., \eqref{LMI11abc} which is usually more difficult to meet for larger $N_0$) whose feasibility requires $\|{\bf b}\|^{2}_{N}$ or $\frac{1}{\alpha}$ to be very small.	This together with \eqref{Hurwitz2} implies that $N$ should be very large. 
\end{remark} \vspace{-0.2cm}

\begin{remark}\label{L0K0design}
Note that for $N_0>1$, it is difficult to find efficient $L_{0}$, $K_{0}$ from \eqref{bL000} (see numerical example in Section \ref{sec4}).
Here for $N_0>1$ we can use the following steps to find more efficient $L_{0}$ and $K_0$:\\
{\bf Step 1}: 
We find $L_0$ from the following inequality:\vspace{-0.3cm}
\begin{equation}\label{L000}
 \begin{array}{ll}
	\setlength{\arraycolsep}{0.1pt}	{\tiny \left[
               \begin{array}{c|cc}
       P_{o}(A_{0}-L_{0}{\bf C}_{0})+(A_{0}-L_{0}{\bf C}_{0})^{\mathrm{T}}P_{o}+2\delta P_{o} & -P_{o}L_{o}  \\\hline
       *&  -\frac{2\delta}{\|{\bf c}\|^{2}_{N}}I
               \end{array}
           \right]  <0}. ~~~~~~~ \vspace{-0.2cm}
	\end{array} 
\end{equation}
The additional terms compared to \eqref{bL000} are from the compensation of infinite-tail term of closed-loop system.\\
{\bf Step 2}: Based on the $L_0$ obtained from \eqref{L000}, we design the controller gain $K_{0}\in \mathbb{R}^{d\times N_{0}}$ from the delay-free case (i.e., $\tau_{u}\equiv 0$ and $\tau_{y}\equiv 0$). In this case, the closed-loop system \eqref{eqXX00} becomes  \vspace{-0.4cm}
\begin{equation*}
	\begin{array}{ll}
	\dot{X}_{0}(t)= F_{0}X_{0}(t) + \mathcal{L}_{0} \zeta(t)  
 +\mathcal{L}_{0}\mathbf{C}_{1}e^{N-N_{0}}(t),  \\
 \dot z_{n}(t)=(-\lambda_{n}+q)z_{n}(t) - B_{n}\mathcal{K}_{0}X_{0}(t), ~n>N.	\vspace{-0.3cm}
	\end{array}
\end{equation*}
We consider vector Lyapunov function \vspace{-0.35cm}
\begin{equation}\label{designFSV}
{\scriptsize \begin{array}{ll}
	V(t)= [V_{0}(t), ~ V_{\mathrm{tail}}(t)]^{\mathrm{T}},\\
	V_{0}(t) =|\hat{z}^{N_{0}}(t)|_{P_z}^{2}+ |e^{N_0}(t)|_{P_e}^{2}+p_{e}|e^{N-N_{0}}(t)|^2,
	 \vspace{-0.3cm}
\end{array} } 
\end{equation}
where $0<P_z, P_e \in \mathbb{R}^{N_{0}\times N_{0}}$, $p_e>0$ and $V_{\mathrm{tail}}(t)$ is defined in \eqref{FSV}. 
By arguments similar to \eqref{Vtail}-\eqref{Hurwitz2}, we have \eqref{Vstable222}  for some $\tilde{D}>0$  provided  \vspace{-0.33cm}  
\begin{equation}\label{designLMI11}
{\scriptsize	\begin{array}{ll}
		\frac{1}{\alpha} K_{0}^{\mathrm{T}}\Lambda_{b}K_{0}< \beta P_z,~~
	{\tiny \left[
               \begin{array}{c|c|c}
             \Phi_{z}   & P_{z}L_{0}{\bf C}_{0}  &   P_{z}L_{0} \\ \hline
                 * & \Phi_{e}  & -P_{e}L_{0}  \\\hline
                 * &*&  -\frac{2\delta_{1}}{\|c\|^2_{N}}I
               \end{array}
           \right]}  <0, \\
    \Phi_{z}=  P_{z}(A_{0}-{\bf B}_{0}K_{0}) 
		+ (A_{0}-{\bf B}_{0}K_{0})^{\mathrm{T}}P_{z} +2\delta P_{z},\\
	\Phi_{e}=	P_{e}(A_{0}-L_{0}{\bf C}_{0})+(A_{0}-L_{0}{\bf C}_{0})^{\mathrm{T}}P_{e} +2\delta P_{e},\\
	2\delta+2\lambda_{N+1}-2q-\alpha>0,\\
	\delta(2\lambda_{N+1}-2q-\alpha)-\beta\delta_{1}>0.	\vspace{-0.33cm}
	\end{array} }
\end{equation}
Let $\beta_{0}=\alpha\beta$, $Q_z=P_{z}^{-1}$ and $Y_{z}=K_{0}Q_{z}$. 
By Schur complement, we find that \eqref{designLMI11} hold iff  \vspace{-0.3cm}  
\begin{equation}\label{designLMI11abc}
{\scriptsize	\begin{array}{ll}
		{\footnotesize \left[
               \begin{array}{ccc}
               -Q_{z}  & Y_{z}^{\mathrm{T}} \\
               *   & - \frac{\beta_0}{\|b\|_{N}^{2}}
               \end{array}
           \right]}<0, ~~   {\tiny \left[
               \begin{array}{c|c|c}
         \tilde{\Phi}_{z}     & L_{0}{\bf C}_0  &   L_{0} \\ \hline
                 * & \Phi_{e}  & -P_{e}L_{0}  \\\hline
                 * &*&  -\frac{2\delta_{1}}{\|c\|^{2}_{N}}I

               \end{array}
           \right]}  <0,\\
   \tilde{\Phi}_{z}=A_{0}Q_{z}+Q_{z}A_{0}^{\mathrm{T}}-B_0 Y_{z}-Y_{z}^{\mathrm{T}}B_{0}^{\mathrm{T}}+2\delta Q_{z},\\
   -2(\lambda_{N+1}-q+\delta) + \alpha <0,\\
	\left[\begin{array}{cc}
	
	 -2\alpha(\lambda_{N+1}-q) + \frac{\delta_{1}}{\delta}\beta_{0} & \alpha \\
	 * & -1
	 \end{array}\right] <0.   \vspace{-0.3cm}  
	\end{array} }
\end{equation}
In particular, \eqref{designLMI11abc} are LMIs that depend on decision variables
$0<Q_z, P_e \in \mathbb{R}^{N_0 \times N_0}$, $Y_z\in\mathbb{R}^{d\times N_{0}}$ and scalars $\alpha, \beta_{0}>0$. If LMIs \eqref{designLMI11abc} hold, the controller gain is given by $K_{0}=Q_{z}^{-1}Y_z$.   
\vspace{-0.2cm}
\end{remark}

\vspace{-0.2cm}
\begin{remark}\label{remark3} 
(Stability analysis via classical Halanay's inequality) 
Consider Lyapunov functional  \vspace{-0.25cm}
\begin{equation}\label{classicalHanalayLy}
	V(t)=V_{0}(t)+V_{\mathrm{tail}}(t)  \vspace{-0.25cm}
\end{equation}
with $V_{0}(t)$ and  $V_{\mathrm{tail}}(t)$ in \eqref{FSV}. To compensate $\zeta(t-\tau_{y})$, the following bound is used for $0<\delta_1 <\delta$: \vspace{-0.35cm}
\begin{equation}\label{halanay}
{\scriptsize  \begin{array}{ll}
 -2\delta_{1}\sup\limits_{t-\tau_{M,y}\leq \theta \leq t}V(\theta) \leq -2\delta_{1}[V_{P}(t-\tau_{y})+ V_{\mathrm{tail}}(t-\tau_{y})] \\
	\overset{\eqref{CSineqaa}}{\leq}-2\delta_{1}  |X_{0}(t)-\nu_{\tau_y}(t)|_{P}^{2} - \frac{2\delta_{1}}{\|{\bf c}\|^{2}_{N}  }|\zeta(t-\tau_{y})|^{2}. \vspace{-0.35cm}
\end{array}	}
\end{equation}
By arguments similar to \eqref{Vtail}, \eqref{dotVX}-\eqref{bLVLV}, \eqref{halanay}, and the following Young inequality for $\alpha_{1}, \alpha_{2}>0$, \vspace{-0.35cm}
\begin{equation}\label{young22}
	{\scriptsize \begin{array}{ll}
		-\sum_{n=N+1}^{\infty}2z_{n}(t){\bf b}^{\mathrm{T}}_{n}\mathcal{K}_{0}X(t-\tau_{u}) \\
		\leq \alpha_{1} \|{\bf b}\|^{2}_{N} |\mathcal{K}_{0}X_0(t)|^{2} + \alpha_{2} \|{\bf b}\|^{2}_{N} |K_{0}\nu_{\tau_{u}}(t)|^{2}\\
		~~~+(\frac{1}{\alpha_{1}}+ \frac{1}{\alpha_{2}} )\sum_{n=N+1}^{\infty} z^{2}_{n}(t), \vspace{-0.35cm}
	\end{array} }
\end{equation}
we have \vspace{-0.35cm}
\begin{equation}\label{cLVLV}
 \begin{array}{ll}
	\dot{V}(t)+ 2\delta V(t) -2\delta_{1}\sup_{t-\tau_{M,y}\leq \theta \leq t}V(\theta)\leq 0 \vspace{-0.3cm}
\end{array}	
\end{equation}
provided \eqref{G1G2} and the following inequalities hold: \vspace{-0.3cm}
\begin{equation}\label{cLMI1}
	\begin{array}{ll}
{\scriptsize	\left[
               \begin{array}{c|cc}
              -\lambda_{N+1}+q+\delta & 1 ~~~~~~~~~~~ 1 \\ \hline
                 * &  \mathrm{diag}\{-2\alpha_{1},-2\alpha_{2}\}
               \end{array}
           \right]<0 },\\
	\Phi_{0}+ \Lambda_{0}^{\mathrm{T}}[\tau_{M,y}^{2}R_{y}  +  \tau_{M,u}^{2}\mathcal{K}^{\mathrm{T}}_{0}R_{u}\mathcal{K}_{0}]\Lambda_{0}  <0, \vspace{-0.3cm}
\end{array}
\end{equation}
where $\Lambda_{0}$ is defined in \eqref{bLMI1a} and \vspace{-0.3cm}
\begin{equation}\label{ccLMI1a}
	\setlength{\arraycolsep}{2pt} {\scriptsize \begin{array}{ll}
		\Phi_{0}={\tiny \left[
               \begin{array}{c|c|c}
             \begin{array}{cc}
                \Omega_{0}  &   P\mathcal{L}_{0}  \\
                * &   -\frac{2\delta_{1}}{\|{\bf c}\|^{2}_{N} }I
               \end{array}   & \begin{array}{cc}
                 \Omega_{1} & \varepsilon_{y}S_{y}  \\
                0 &  0
               \end{array}
              &   \begin{array}{cc}
                 P\mathcal{B}_{0} +\varepsilon_{u}\mathcal{K}_{0}^{\mathrm{T}}S_{u}   & \varepsilon_{u}\mathcal{K}_{0}^{\mathrm{T}} S_{u} \\
                0 &  0
               \end{array}  \\ \hline
                 * & \Omega_{y}  & 0 \\\hline
                 * &*&  \Omega_{u}
               \end{array}
           \right]}, \\
	 \Omega_{0}= PF_{0}+F_{0}^{\mathrm{T}}P+2(\delta-\delta_{1} )P +(1-\varepsilon_{u})\mathcal{K}^{\mathrm{T}}_{0}S_{u}\mathcal{K}_{0} \\
	 ~~~~~~+(1-\varepsilon_{y})S_{y} +  \alpha_{1} \|{\bf b}\|^{2}_{N}  \mathcal{K}_{0}^{\mathrm{T}} \mathcal{K}_{0},  \\	
	 \Omega_{1}= 2\delta_{1}P-P\mathcal{L}_{0}\mathcal{C}_0 +\varepsilon_{y}S_{y} ,\\
     \Omega_{y} = {\tiny \left[
               \begin{array}{cc}
               -2\delta_{1}P-\varepsilon_{y}(S_{y}+R_{y}) &~~ -\varepsilon_{y}(S_{y}+G_{y})  \\
                 * &  ~~-\varepsilon_{y}(S_{y}+R_{y})
               \end{array}
           \right] }, \\
    \Omega_{u} ={\tiny \left[
               \begin{array}{cc}
               \alpha_{2} \|{\bf b}\|^{2}_{N} I -\varepsilon_{u}(S_{u}+R_{u}) & ~~-\varepsilon_{u}(S_{u}+G_{u})  \\
                 * &  ~~-\varepsilon_{u}(S_{u}+R_{u})
               \end{array}
           \right] }.  \vspace{-0.3cm}
	\end{array} }
\end{equation}
Then classical Halanay's inequality (see P. 138 in \cite{fridman2014introduction}) and \eqref{cLVLV} imply \eqref{Vstable222}, where $\delta_{0}>0$ is the unique solution of $\delta_{0}=\delta-\delta_{1}\mathrm{e}^{2\delta_{0}\tau_{M,y}}$. 
\end{remark}


\vspace{-0.3cm}
\section{Non-local actuation and boundary measurement}\label{sec3}\vspace{-0.35cm}
Consider system \eqref{pde1aab} with ${\bf b}\in (H^{1}(\Omega))^d$, ${\bf b}(x)=0$ for $x\in \Gamma_{D}$. Let $N_{0}\in \mathbb{N}$ satisfy \eqref{eq4},  $N\geq N_{0}$, and $d$ be the maximum of the geometric multiplicities of $\lambda_{n}$, $n=1,\dots, N_{0}$. 
We assume the following delayed boundary measurement: \vspace{-0.35cm}
\begin{equation}\label{measure155}
{\scriptsize	\begin{array}{ll}
		y(t)= \int_{\Gamma_{N}} {\bf c}(x) z(x, t-\tau_{y}))\mathrm{d}x, ~t-\tau_{y} \geq 0,\\
		y(t)=0, t-\tau_{y}<0,~{\bf c}=[c_1, \dots, c_{d}]^{\mathrm{T}} \in (L^{2}(\Gamma_{N}))^d. \vspace{-0.3cm}
	\end{array} }
\end{equation}
Note that \eqref{measure155} is actually a weighted averaged boundary measurement with ${\bf c}$ representing the weighted coefficient. 
We present the solution to \eqref{pde1aab} as \eqref{eq2a00} with $z_{n}$ satisfying \eqref{obeq33aa00}.  We construct a $N$-dimensional observer of the form \eqref{eqaa100}, where  $\hat{z}_{n}(t)$ $(1\leq n\leq N)$ satisfy   \vspace{-0.4cm}
\begin{equation}\label{observer22300cc}
  \begin{array}{ll}
  \dot{\hat{z}}_{n}(t)=(-\lambda_{n}+q)\hat{z}_{n}(t)+b_{n}u(t)\\
    ~~~~~~~-l_{n}[\sum_{i=1}^{N}{\bf c}_{i}\hat{z}_{i}(t-\tau_{y})  -y(t) ], ~t> 0, \\
  \hat{z}_{n}(0)=0, ~t\leq 0,~~ 
  {\bf c}_{i}=\int_{\Gamma_{N}} {\bf c}(x)\phi_{i}(x)\mathrm{d}x, \vspace{-0.35cm}
  \end{array}
\end{equation}
with $y(t)$ in \eqref{measure155} and observer gains 
$\{l_{n}\}_{n=1}^{N}$, $l_{n}\in\mathbb{R}^{1\times d}$. 
In this section, all notations are the same as in Sec. \ref{sec2} except of 
${\bf c}_{n}$ which are defined in \eqref{observer22300cc}. Let ${\bf B}_{0}$ and ${\bf C}_{0}$ satisfy Assumption \ref{assump1}. From Lemma \ref{proposition2.1}, we let  $L_{0}=\mathrm{col}\{l_{1},\dots , l_{N_{0}}\}\in\mathbb{R}^{N_{0}\times d}$  satisfy \eqref{bL00011}. Define $u(t)$ in \eqref{eq600} with $K_{0}\in \mathbb{R}^{d\times N_{0}}$ satisfying \eqref{bL00022}.
By \eqref{eq600}, \eqref{tail100aa}, \eqref{error00}, \eqref{observer22300cc}, and $X_0(t)$ defined in \eqref{symbol00}, we obtain the closed-loop system \eqref{eqXX00}.

\vspace{-0.1cm}
 Note that we need \eqref{CSineqaa} to compensate $\zeta(t-\tau_y)$ in \eqref{eqXX0011} by Halanay inequality. However, differently from the non-local measurement where $\sum_{n=N+1}^{\infty}|{\bf c}_{n}|^{2}<\infty$, for the boundary measurement with ${\bf c}_{n}$ defined in \eqref{observer22300cc}, we do not have this property. Here we assume  \vspace{-0.35cm}
\begin{equation}\label{eq3.5}
	\begin{array}{ll}
		\sum_{n=N+1}^{\infty}\frac{|{\bf c}_{n}|^2}{\lambda_{n}}\leq \varrho_{N}\leq \varrho, \vspace{-0.35cm}
	\end{array}
\end{equation}
for some $\varrho_{N}>0$, where $\varrho>0$ is independent of $N$. For $\zeta(t)$ defined in \eqref{tail100aa}, by Cauchy-Schwarz inequality, we have \vspace{-0.35cm}
\begin{equation}\label{CSineq}
\begin{array}{ll}
	|\zeta(t)|^{2}
	\leq \sum_{n=N+1}^{\infty}\frac{|{\bf c}_{n}|^2}{\lambda_{n}}\sum_{n=N+1}^{\infty}\lambda_{n}z^{2}_{n}(t)\\
	\overset{\eqref{eq3.5}}\leq \varrho_{N}\sum_{n=N+1}^{\infty}\lambda_{n}z^{2}_{n}(t). \vspace{-0.35cm}
\end{array}	
\end{equation} 
\vspace{-0.3cm}
\begin{remark}\label{rmk3.1}
	Note that \eqref{eq3.5} holds for rectangular domain $\Omega=(0,a_{1})\times (0,a_{2})$ with the following boundary \vspace{-0.3cm}
\begin{equation}\label{boundary}
{\scriptsize	\begin{array}{ll}
		\Gamma=\Gamma_{D}\cup \Gamma_{N},~~\Gamma_{N}= \{(x_{1},0), ~x_{1}\in(0,a_{1})\},\\
		\Gamma_{D}= \{(0,x_{2}),x_{2}\in[0,a_{2}]\} \cup \{(a_{1},x_{2}),x_{2}\in[0,a_{2}]\} \\
		~~~~~~~~~\cup\{(x_{1},a_{2}), x_{1}\in [0,a_{1}]\}. \vspace{-0.3cm}
	\end{array} }
\end{equation}
The eigenvalues and corresponding eigenfunctions of $\mathcal{A}$ (see \eqref{operatorA}) are given by: \vspace{-0.3cm}
\begin{equation}\label{eigenvalue}
\begin{array}{ll}
	\lambda_{m,k}=\pi^{2}[\frac{m^2}{a_{1}^{2}} +\frac{(k-\frac{1}{2})^2}{a_{2}^2}], ~~m,k\in \mathbb{N}, \\
	 \phi_{m,k}(x_{1},x_{2})=\frac{2}{\sqrt{a_{1}a_{2}}}\sin(\frac{m\pi x_{1}}{a_{1}})\cos(\frac{(k-\frac{1}{2})\pi x_{2}}{a_{2}}). \vspace{-0.25cm}
\end{array}	
\end{equation}
We reorder the eigenvalues \eqref{eigenvalue} to form a non-decreasing  sequence \eqref{eigenvalue2} and denote the corresponding eigenfunctions as $\{\phi_{n}\}_{n=1}^{\infty}$.	 Let the corresponding relationship between \eqref{eigenvalue} and \eqref{eigenvalue2} be $n\sim (m,k)$. We have $|{\bf c}_{n}|^{2}=|{\bf c}_{m,k}|^{2}=\sum_{j=1}^{d}|\int_{0}^{a_{1}}c_{j}(x_{1})\phi_{m,k}(x_{1},0)\mathrm{d}x_{1}|^{2}=\frac{2}{a_{2}}\sum_{j=1}^{d}c_{j,m}^{2}$ where $c_{j,m}=\int_{0}^{a_{1}} c_{j}(x_{1})\cdot \frac{\sqrt{2}}{\sqrt{a_{1}}}\sin(\frac{m \pi x_{1}}{a_{1}})\mathrm{d}x_{1}$ satisfying 
 $\|c_{j}\|^{2}_{L^{2}(\Gamma_{N})}=\sum_{m=1}^{\infty}c_{j,m}^{2}$.  Therefore, we have \vspace{-0.3cm}
\begin{equation}\label{CSineq22}
	\begin{array}{ll}
	\sum_{n=1}^{\infty}\frac{|{\bf c}_{n}|^2}{\lambda_{n}} = \frac{2}{a_{2}} \sum_{j=1}^{d}\sum_{m,k=1}^{\infty}   \frac{ c_{j,m}^{2}}{ \lambda_{m,k} } \\
		\leq \sum_{j=1}^{d}\sum_{m=1}^{\infty} c_{j,m}^{2}\sum_{k=1}^{\infty}\frac{2a_{2}}{(k-\frac{1}{2})^{2}\pi^{2}} \\
	=a_{2}\sum_{j=1}^{d}\|c_{j}\|^{2}_{L^{2}(0,a_{1})}=:\varrho,  \vspace{-0.3cm}
	\end{array}
\end{equation}
where $\sum_{k=1}^{\infty}\frac{1}{(2k-1)^{2}}=\frac{\pi^{2}}{8}$ \footnote{The proof can be found on this website: \href{https://daviddeley.com/pendulum/page17proof.htm}{https://daviddeley.com/pendulum/page17proof.htm}.} 
is used and $\varrho$ is independent of $N$. From \eqref{CSineq22} it follows\vspace{-0.35cm}
\begin{equation}\label{CSineq22abc}
	\begin{array}{ll}
		\sum_{n=N+1}^{\infty}\frac{|{\bf c}_{n}|^2}{\lambda_{n}} 
		\leq \varrho - \sum_{n=1}^{N}\frac{|{\bf c}_{n}|^2}{\lambda_{n}}=:\varrho_{N}.   \vspace{-0.3cm}
	\end{array}
\end{equation}
\end{remark}

\vspace{-0.2cm}
 Taking into account \eqref{CSineq}, for exponential $H^{1}$-stability we consider the vector Lyapunov functional \eqref{FSV} with $V_{\mathrm{tail}}(t)$ therein replaced by \vspace{-0.35cm}
\begin{equation}\label{taiVV}
	\begin{array}{ll}
		V_{\mathrm{tail}}(t)=\sum_{n=N+1}^{\infty}\lambda_{n}z^{2}_{n}(t). \vspace{-0.3cm}
	\end{array}
\end{equation}
Differentiation of $V_{\mathrm{tail}}(t)$ in \eqref{taiVV} along \eqref{eqXX0022} gives \vspace{-0.35cm}
\begin{equation}\label{VtailboundaryM}
	\begin{array}{ll}
		\dot{V}_{\mathrm{tail}}(t)= \sum_{n=N+1}^{\infty}2(-\lambda_{n}+q)\lambda_{n}z^{2}_{n}(t) \\
		~~~-\sum_{n=N+1}^{\infty}2\lambda_{n}z_{n}(t){\bf b}^{\mathrm{T}}_{n}\mathcal{K}_{0}X(t-\tau_{u})\\
		\leq \sum_{n=N+1}^{\infty}2(-\lambda_{n}+q+ \alpha )\lambda_{n}z^{2}_{n}(t) \\
		~~~+ \frac{1}{\alpha} \|\nabla {\bf b}\|^{2}_{N}|\mathcal{K}_{0}X(t-\tau_{u})|^{2} \vspace{-0.45cm}
	\end{array}
\end{equation} 
for some $\alpha>0$, where $\|\nabla {\bf b}\|^{2}_{N}=\sum_{j=1}^{d}\|\nabla  b_j\|^{2}_{N}\overset{\eqref{inequality00}} =\sum_{j=1}^{d}\sum_{n=N+1}^{\infty}\lambda_{n}\langle b_{j},\phi_{n}\rangle^{2}$. 
By arguments similar to \eqref{Vtail}-\eqref{Hurwitz2} and using  \eqref{CSineq}, \eqref{VtailboundaryM}, we obtain \vspace{-0.3cm}
\begin{equation}\label{VstableH11}
	 \|z(\cdot,t)\|^{2}_{H^{1}}+\|z(\cdot,t)-\hat{z}(\cdot,t) \|^{2}_{H^{1}}\leq \tilde{D}\mathrm{e}^{-\delta_{0}t},~t\geq 0 \vspace{-0.3cm}
\end{equation}
for some $\tilde{D}>0$ and $\delta_{0}>0$ provided LMIs \eqref{LMI11abc} (where $\|{\bf b}\|^{2}_{N}$ is changed to $\|\nabla {\bf b}\|^{2}_{N}$), 
\eqref{G1G2}, \eqref{bLMI1} with  $\Phi_{0}$ (where $\|{\bf c}\|^{2}_{N}$ is changed to $\varrho_{N}$) and $\Lambda_{0}$ given in \eqref{bLMI1a}, and \eqref{Hurwitz2} hold. 
The asymptotic feasibility of above LMIs for large enough $N$ and small enough $\tau_{M,y}, \tau_{M,u}>0$ can be obtained by arguments similar to Theorem \ref{thm1}. Summarizing, we arrive at: \vspace{-0.18cm}
\begin{theorem}\label{thm3}
Consider \eqref{pde1aab} with control law \eqref{eq600}  
where ${\bf b}\in (H^{1}(\Omega))^d$, ${\bf b}(x)=0$ for $x\in \Gamma_{D}$, measurement \eqref{measure155}, and $z_{0}\in \mathcal{D}(\mathcal{A})$.
Given $\delta,\delta_{1}>0$, let $N_{0}\in\mathbb{N}$ satisfy \eqref{eq4} and $N\in\mathbb{N}$ satisfy $N\geq N_{0}$. 
 Let Assumption \ref{assump1} and \eqref{eq3.5} hold and $L_{0}$, $K_{0}$ be obtained from \eqref{bL000}. 
Given $\tau_{M,y}, \tau_{M,u}>0$, let there exist $0<P, S_{y}, R_{y} \in \mathbb{R}^{2N_{0}\times 2N_{0}}$, $0<S_{u}, R_{u}\in\mathbb{R}^{d\times d}$, scalars $\alpha, \beta_{0}>0$, $G_{y}\in \mathbb{R}^{2N_{0}\times 2N_{0}}$ and $G_{u}\in \mathbb{R}^{d\times d}$  
such that LMIs \eqref{LMI11abc} (where $\|{\bf b}\|^{2}_{N}$ is changed to $\|\nabla  {\bf b}\|^{2}_{N}$), 
\eqref{G1G2}, \eqref{bLMI1} with  $\Phi_{0}$ (where $\|{\bf c}\|^{2}_{N}$ is changed to $\varrho_{N}$) and $\Lambda_0$ given in \eqref{bLMI1a}, and \eqref{Hurwitz2} hold.
Then the solution $z(x,t)$ to \eqref{pde1aab} subject to the control law \eqref{observer22300}, \eqref{eq600} and the corresponding observer $\hat{z}(x,t)$ given by \eqref{eqaa100} satisfy \eqref{VstableH11}.
Moreover, the above LMIs always hold for large enough $N$ and small enough $\tau_{M,y},\tau_{M,u}>0$.
\end{theorem}

\vspace{-0.25cm}
\begin{remark}\label{rem4}
	(Stability analysis via classical Halanay's inequality)
Consider Lyapunov functional \eqref{classicalHanalayLy} with $V_{0}(t)$ in \eqref{FSV} and  $V_{\mathrm{tail}}(t)$ in \eqref{taiVV}. By arguments similar to \eqref{Vtail}-\eqref{Hurwitz2} and using following bound for $0<\delta_1 <\delta$: \vspace{-0.3cm}
\begin{equation*}
{\scriptsize  \begin{array}{ll}
 -2\delta_{1}\sup\limits_{t-\tau_{M,y}\leq \theta \leq t}V(\theta) \leq -2\delta_{1}[V_{P}(t-\tau_{y})+ V_{\mathrm{tail}}(t-\tau_{y})] \\
	\overset{\eqref{CSineq}}{\leq}-2\delta_{1}  |X_{0}(t)-\nu_{\tau_y}(t)|_{P}^{2} - \frac{2\delta_{1}}{ \varrho_{N}  }|\zeta(t-\tau_{y})|^{2}, \vspace{-0.1cm}
\end{array}	}
\end{equation*}
and the following Young inequality for $\alpha_{1}, \alpha_{2}>0$, \vspace{-0.3cm}
\begin{equation*}
	{\scriptsize \begin{array}{ll}
		-\sum_{n=N+1}^{\infty}2\lambda_{n}z_{n}(t){\bf b}^{\mathrm{T}}_{n}\mathcal{K}_{0}X(t-\tau_{u}) \\
		\leq \alpha_{1} \|\nabla {\bf b}\|^{2}_{N} |\mathcal{K}_{0}X_0(t)|^{2} + \alpha_{2} \|\nabla  {\bf b}\|^{2}_{N} |K_{0}\nu_{\tau_{u}}(t)|^{2}\\
		~~~+(\frac{1}{\alpha_{1}}+ \frac{1}{\alpha_{2}} )\sum_{n=N+1}^{\infty} \lambda_{n}z^{2}_{n}(t), \vspace{-0.3cm}
	\end{array} }
\end{equation*}
we obtain 
\eqref{VstableH11} provided \eqref{G1G2} and \eqref{cLMI1} hold with 
 $\Lambda_0$ in \eqref{bLMI1a} and $\Phi_{0}$, $\Omega_{y}$, $\Omega_{u}$ in \eqref{ccLMI1a} (where $\|{\bf b}\|^{2}_{N}$ and $\|{\bf c}\|^{2}_{N}$ are changed to $\|\nabla {\bf b}\|^{2}_{N}$ and $\varrho_{N}$, respectively).
\end{remark}
\vspace{-0.3cm}

\section{Boundary actuation and non-local measurement}\label{sec4}\vspace{-0.3cm}
Consider the delayed Neumann actuation \vspace{-0.3cm}
\begin{equation}\label{pde1aab2}
\begin{array}{ll}
	z_{t}(x,t)=\Delta z(x,t)+qz(x,t) , ~\mathrm{in}~ \Omega \times (0,\infty),\\
	z(x,t)=0,~~\mathrm{on}~ \Gamma_{D}\times (0,\infty), \\
	\frac{\partial z}{\partial {\bf n}}(x,t)={\bf b}^{\mathrm{T}}(x)u(t-\tau_{u}),~~\mathrm{on}~ \Gamma_{N}\times (0,\infty), \\
	z(x,0)=z_{0}(x), ~~x\in \Omega, \vspace{-0.3cm}
	\end{array}
\end{equation}
where ${\bf b}=[b_{1},\dots,b_{d}]^{\mathrm{T}}\in (L^{2}(\Gamma_{N}))^{d}$ and $u(t)=[u_{1}(t),\dots,u_{d}(t)]^{\mathrm{T}}$ is the control input to be designed. Let $N_{0}\in \mathbb{N}$ satisfy \eqref{eq4} and $N\geq N_{0}$. Let $d$ be the maximum of the geometric multiplicities of $\lambda_{n}$, $n=1,\dots, N_{0}$.
We consider 
the delayed non-local measurement \eqref{measure133} with ${\bf c}\in (L^{2}(\Omega))^d$. 
We present the solution to \eqref{pde1aab2} as \eqref{eq2a00} and obtain \eqref{obeq33aa00} with \vspace{-0.4cm}
\begin{equation}\label{controllerbn}
	\begin{array}{ll}
	{\bf b}_{n}= \int_{\Gamma_{D}} {\bf b}(x)\phi_{n}(x)\mathrm{d}x \vspace{-0.35cm}
	\end{array}
\end{equation}

\vspace{-0.2cm}
In this section, all notations are the same as in Sec. \ref{sec2} except of  ${\bf b}_{n}$ that are defined by \eqref{controllerbn}. 
We construct a $N$-dimensional observer of the form \eqref{eqaa100}, where $\hat{z}_{n}(t)$ satisfy \eqref{observer22300}. Let ${\bf B}_{0}$ and ${\bf C}_{0}$ satisfy Assumption \ref{assump1}.  
From Lemma \ref{proposition2.1}, let  $L_{0}=\mathrm{col}\{l_{1},\dots , l_{N_{0}}\}\in\mathbb{R}^{N_{0}\times d}$  satisfy \eqref{bL00011}. Define $u(t)$ in \eqref{eq600} with $K_{0}\in \mathbb{R}^{d\times N_{0}}$ satisfying \eqref{bL00022}.

\vspace{-0.15cm}
For the well-posedness of closed-loop system \eqref{pde1aab2} and  \eqref{observer22300}, with control input \eqref{eq600}, we introduce the change of variables \vspace{-0.3cm}
\begin{equation}\label{changeofvariable}
		w(x,t)= z(x,t)-{\bf r}^{\mathrm{T}}(x)u(t-\tau_{u}),  \vspace{-0.3cm}
\end{equation}
where ${\bf r}(x)=[r_{1}(x),\dots, r_{d}(x)]^{\mathrm{T}}$ with $r_{j}(x)$, $j=1,\dots,d$ being the solution to the following Laplace equation: \vspace{-0.3cm}
\begin{equation}\label{rrrr}
	\begin{array}{ll}
		\Delta r_{j}(x) = 0, ~~x\in \Omega,\\
		r_{j}(x)=0,~x\in\Gamma_{D}, ~~\frac{\partial r_{j}}{\partial {\bf n}}(x)=b_{j}(x), ~x\in \Gamma_{N}. \vspace{-0.3cm}
 	\end{array}
\end{equation}
Since $b_{j}\in L^{2}(\Gamma_{N})$, from \cite[Lemma 2.1]{feng2022boundary} we have $r_{j}\in L^{2}(\Omega)$. By \eqref{pde1aab2}, \eqref{changeofvariable}, and \eqref{rrrr}, we get the equivalent evolution equation: \vspace{-0.3cm}
\begin{equation}\label{equivalentPDE}
	\begin{array}{ll}
		\dot{w}(t)+ \mathcal{A}w(t)=qw(t)-{\bf r}^{\mathrm{T}}(\cdot)\dot{u}(t-\tau_u)(1-\dot{\tau}_{u}) \\
		~~~~~~~~~~~~~~~~~~~~+ q{\bf r}^{\mathrm{T}}(\cdot) u(t-\tau_{u}),\\
		w(0)=z(\cdot,0). \vspace{-0.35cm}
	\end{array}
\end{equation}
Define the state $\xi(t)=\mathrm{col}\{w(t), \hat{z}^{N}(t)\}$, where $\hat{z}^{N}(t)=[\hat{z}_{1}(t), \dots, \hat{z}_{N}(t)]^{\mathrm{T}}$. By \eqref{observer22300}, \eqref{eq600}, and \eqref{equivalentPDE}, we present the closed-loop system as \vspace{-0.35cm}
\begin{equation}\label{abstractsystem22}
\setlength{\arraycolsep}{1.5pt}	
{\scriptsize \begin{array}{ll}
		\frac{\mathrm{d}}{\mathrm{d}t}\xi(t)+ \mathrm{diag}\{\mathcal{A},  \mathcal{A}_{0}\} \xi(t)
		= { \tiny \left[\begin{array}{ccc}
		qw(t)+f_{1}(t-\tau_{u}) \\
		 f_{2}(t-\tau_{u}) + f_{3}(t-\tau_{y}) \\
		 0_{(N-N_{0})\times 1} \end{array} \right]},\\
	f_{3}(t)=-L_{0}[{\bf C}\hat{z}^{N}(t) -\langle {\bf c},w(\cdot,t) \rangle  + \langle {\bf c},  {\bf r}^{\mathrm{T}}(\cdot) K_{0}\hat{z}^{N_{0}}(t-\tau_{u})    \rangle],\\
	f_{1}(t)= {\bf r}^{\mathrm{T}}(\cdot)(1-\dot{\tau}_{u})K_{0}[ A_{0}\hat{z}^{N_{0}}(t)+ f_{3}(t-\tau_{y})  \\
	~~~~~~~ ~~~-B_{0}K_{0}\hat{z}^{N_{0}}(t-\tau_{u}) ] - q{\bf r}^{\mathrm{T}}(\cdot)\hat{z}^{N_{0}}(t),   \vspace{-0.3cm}
	\end{array} }
\end{equation}
where $\mathcal{A}_{0}$, ${\bf C}$, and $f_{2}(t)$ are defined in \eqref{abstractsystem}. By arguments similar to the well-posedness in Sec. \ref{sec2},
we obtain that \eqref{abstractsystem22} has a unique solution satisfying  \eqref{regularity}. From \eqref{changeofvariable}, it follows \eqref{pde1aab2}, subject to \eqref{observer22300}, \eqref{eq600},  has a unique classical solution such that $z\in C([0,\infty),L^{2}(\Omega))\cap C^{1}((0,\infty),L^{2}(\Omega))$ and $z(\cdot,t)\in H^{2}(\Omega)$ with $z(x,t)=0$, $x\in\Gamma_{D}$ and $\frac{\partial}{\partial {\bf n}}z(x,t)={\bf b}^{\mathrm{T}}(x)u(t-\tau_{u})$, $x\in \Gamma_{N}$, for $t\in[0,\infty)$.

With notations \eqref{symbol00}, the closed-loop system has a form: \vspace{-0.3cm}
\begin{equation}\label{eqXX11}
\setlength{\arraycolsep}{1.5pt} {\scriptsize \begin{array}{ll}
\dot{X}_{0}(t)= F_{0}X_{0}(t) - \mathcal{L}_{0}\mathcal{C}\nu_{\tau_{y}}(t) + \mathcal{B}\mathcal{K}_{0}\nu_{\tau_{u}}(t)  + \mathcal{L}_{0} \zeta(t-\tau_{y}),  \\
 \dot z_{n}(t)=(-\lambda_{n}+q)z_{n}(t) - {\bf b}^{\mathrm{T}}_{n}\mathcal{K}_{0}X_{0}(t-\tau_{u}), ~ n>N.  \vspace{-0.35cm}
\end{array} }
\end{equation}
For non-local actuation case in Sec. \ref{sec2}, we employ Young's inequality \eqref{byoung} to split the finite- and infinite-dimensional parts, where $\sum_{n=N+1}^{\infty}|{\bf b}_{n}|^2<\infty$ is used.  However, for the boundary actuation with ${\bf b}_{n}$ defined in \eqref{controllerbn}, we do not have such property. 
Here we assume \vspace{-0.3cm}
\begin{equation}\label{eq4.2}
	\begin{array}{ll}
		\sum_{n=N+1}^{\infty}\frac{|{\bf b}_{n}|^2}{\lambda_{n}}\leq \rho_{N}\leq \rho, \vspace{-0.3cm}
	\end{array}
\end{equation}
for some $\rho_{N}>0$, where $\rho>0$ is independent of $N$. 
Then we use the following Young inequality for $\alpha>0$: \vspace{-0.3cm}
\begin{equation}\label{Yang3}
{\scriptsize	\begin{array}{ll}
		-\sum_{n=N+1}2z_{n}(t){\bf b}_{n}^{\mathrm{T}}\mathcal{K}_{0}X_{0}(t-\tau_{u}) \\
		\leq {1 \over \alpha} \sum^{\infty}_{n=N+1}\frac{|{\bf b}_{n}|^{2} }{\lambda_{n}} |\mathcal{K}_{0}X_{0}(t-\tau_{u}) |^{2} +\sum_{n= N+1}^{\infty}\alpha \lambda_{n}  z^{2}_{n}(t) \\
	\overset{\eqref{eq4.2}}\leq {\rho_{N} \over \alpha } |\mathcal{K}_{0}X_{0}(t-\tau_{u})|^{2} +\sum_{n=N+1}^{\infty}\alpha\lambda_{n}  z^{2}_{n}(t). \vspace{-0.4cm}
	\end{array} }
\end{equation}  
\vspace{-0.6cm}
\begin{remark}
Note that \eqref{eq4.2} holds for rectangular domain. Consider the rectangular domain introduced in Remark  \ref{rmk3.1}. 
Similar to estimates \eqref{CSineq22} and \eqref{CSineq22abc}, we have $ \sum_{n =N+1}^{\infty}\frac{|{\bf b}_{n}|^{2}}{\lambda_{n}}
	\leq \rho -  \sum_{n=1}^{N}\frac{|{\bf b}_{n}|^{2}}{\lambda_{n}}=: \rho_{N}$
with $\rho=a_{2}\sum_{j=1}^{d}\|b_{j}\|^{2}_{L^{2}(0,a_1)} $ which is independent of $N$.
\end{remark}

\vspace{-0.1cm}
According to \eqref{Yang3}, we consider the following Cauchy-Schwarz inequality:\vspace{-0.3cm}
\begin{equation}\label{Cauchy-Schwarz22}
	\begin{array}{ll}
		|\zeta(t)|^{2}\leq \sum_{n=N+1}^{\infty}\frac{|{\bf c}_{n}|^{2}}{\lambda_{n}} \sum_{n=N+1}^{\infty}\lambda_{n}z^{2}_{n}(t) \\
		\leq \frac{\|{\bf c} \|^{2}_{N}}{\lambda_{N} } \sum_{n=N+1}^{\infty}\lambda_{n}z^{2}_{n}(t), \vspace{-0.3cm}
	\end{array}
\end{equation}
where $\|{\bf c}\|^{2}_{N}$ is defined in \eqref{CSineqaa}.
Consider the vector Lyapunov functional \eqref{FSV} with $V_{\mathrm{tail}}(t)$ therein replaced by \eqref{taiVV}. 
By arguments similar to \eqref{VXVXV}-\eqref{Hurwitz2}, \eqref{VtailboundaryM}, and using \eqref{Yang3} and \eqref{Cauchy-Schwarz22}, we conclude that the solutions to \eqref{pde1aab2}, \eqref{observer22300}, \eqref{eq600} satisfy \eqref{VstableH11} for some $\tilde{D}>0$ and $\delta_{0}>0$ provided
 \eqref{G1G2}, \eqref{bLMI1ab} with $\Phi_{0}$, $\Lambda_{0}$ in \eqref{bLMI1a} (where $\|{\bf c}\|_{N}^{2}$ is changed to $\frac{1}{\lambda_{N}} \|{\bf c}\|^{2}_{N}$), and the following inequalities hold: \vspace{-0.2cm}
\begin{equation}\label{LMI11abcabc}
	\begin{array}{ll}
		{\tiny \left[
               \begin{array}{ccc}
               -P  &  \mathcal{K}_{0}^{\mathrm{T}} \\
               *   & -\frac{\beta_{0}}{ \rho_{N} }I
               \end{array}
           \right]}<0,
          { \tiny \left[
               \begin{array}{ccc}
              -2\alpha(\lambda_{N+1}-q) +\frac{\delta_{1}}{\delta}\beta_{0} &  \alpha \\
               *    &  - 1
               \end{array}
           \right]} <0. ~~~~~~~~~~~~~~~~~~\vspace{-0.2cm}
	\end{array}
\end{equation}
The asymptotic feasibility of above LMIs for large enough $N$ and small enough $\tau_{M,y}, \tau_{M,u}>0$ can be obtained by arguments similar to Theorem \ref{thm1}. Summarizing, we have: \vspace{-0.3cm}
\begin{theorem}\label{thm2}
Consider \eqref{pde1aab2} with control law \eqref{eq600} and delayed non-local measurement \eqref{measure133}.
Given $\delta>0$, let $N_{0}\in\mathbb{N}$ satisfy \eqref{eq4} and $N\in\mathbb{N}$ satisfy $N\geq N_{0}$.
Let Assumption \ref{assump1} hold and $L_{0}\in\mathbb{R}^{N_{0}\times d}$, $K_{0}\in \mathbb{R}^{d\times N_{0}}$ be obtained from \eqref{bL000}.
Given $\tau_{M,y}, \tau_{M,u}>0$, let there exist $0<P, S_{y}, R_{y} \in \mathbb{R}^{2N_{0} \times 2N_{0}}$, $0<S_{u}, R_{u}\in\mathbb{R}^{d\times d}$, $G_{y}\in \mathbb{R}^{2N_{0} \times 2N_{0}}$ and $G_{u}\in \mathbb{R}^{d\times d}$, scalars $\alpha, \beta_0>0$  such that LMIs \eqref{G1G2}, \eqref{bLMI1} with  $\Phi_{0}$ and $\Lambda_{0}$ given in \eqref{bLMI1a} (where $\|{\bf c}\|^{2}_{N}$ is changed to  $\|{\bf c}\|^{2}_{N}/\lambda_{N}$) and \eqref{LMI11abcabc} hold.
Then the solution $z(x,t)$ to \eqref{pde1aab2} subject to the control law \eqref{observer22300}, \eqref{eq600} and the corresponding observer $\hat{z}(x,t)$ given by \eqref{eqaa100} satisfy \eqref{VstableH11} for some $\tilde{D}>0$ and $\delta_{0}>0$.
Moreover, the above inequalities always hold for large enough $N$ and small enough $\tau_{M,y},\tau_{M,u}>0$.
\end{theorem}
 \vspace{-0.2cm}
\begin{remark}\label{rem5}
	(Stability analysis via classical Halanay's inequality)
Consider Lyapunov functional \eqref{classicalHanalayLy} with $V_{0}(t)$ in \eqref{FSV} and  $V_{\mathrm{tail}}(t)$ in \eqref{taiVV}. By arguments similar to \eqref{Vtail}-\eqref{Hurwitz2} and using following bound for $0<\delta_1 <\delta$: \vspace{-0.3cm}
\begin{equation*}
{\scriptsize  \begin{array}{ll}
 -2\delta_{1}\sup\limits_{t-\tau_{M,y}\leq \theta \leq t}V(\theta) \leq -2\delta_{1}[V_{P}(t-\tau_{y})+ V_{\mathrm{tail}}(t-\tau_{y})] \\
	\overset{\eqref{Cauchy-Schwarz22}}{\leq}-2\delta_{1}  |X_{0}(t)-\nu_{\tau_y}(t)|_{P}^{2} - \frac{2\delta_{1}\lambda_{N}}{ \|{\bf c}\|_{N}  }|\zeta(t-\tau_{y})|^{2}, 
\end{array}	}
\end{equation*}
and the following Young inequality for $\alpha_{1}, \alpha_{2}>0$, \vspace{-0.3cm}
\begin{equation*}
	{\scriptsize \begin{array}{ll}
		-\sum_{n=N+1}^{\infty}2\lambda_{n}z_{n}(t){\bf b}^{\mathrm{T}}_{n}\mathcal{K}_{0}X(t-\tau_{u}) \\
	\overset{\eqref{eq4.2}}	\leq \alpha_{1}\rho_{N} |\mathcal{K}_{0}X_0(t)|^{2} + \alpha_{2}\rho_{N} |K_{0}\nu_{\tau_{u}}(t)|^{2}\\
		~~~+(\frac{1}{\alpha_{1}}+ \frac{1}{\alpha_{2}} )\sum_{n=N+1}^{\infty} \lambda_{n}z^{2}_{n}(t), \vspace{-0.3cm}
	\end{array} }
\end{equation*}
we obtain 
\eqref{VstableH11} provided \eqref{G1G2} and \eqref{cLMI1} hold with 
 $\Lambda_0$ in \eqref{bLMI1a} and $\Phi_{0}$, $\Omega_{y}$, $\Omega_{u}$ in \eqref{ccLMI1a} (where $\|{\bf b}\|^{2}_{N}$ and $\|{\bf c}\|^{2}_{N}$ are changed to $\rho_{N}$ and $\|{\bf c}\|^{2}_{N}/\lambda_{N}$, respectively).
\end{remark}
 \vspace{-0.2cm}



\vspace{-0.1cm}
\section{Numerical examples}\label{sec5} \vspace{-0.3cm}
In this section, we consider a rectangular domain $\Omega=(0,a_{1})\times (0,a_2)$ with $a_1=\frac{4\sqrt{3}}{3}$, $a_2=\frac{4\sqrt{3}}{3}$ and boundary \eqref{boundary}. 
We consider $q=3$ which results in an unstable open-loop system with 1 unstable mode (in this case, $N_0=1$ and $d=1$) and  $q=8.1$ which results in an unstable open-loop system with 3 unstable modes with $\lambda_{1}<\lambda_2 =\lambda_3$ (in this case, $N_0=3$ and $d=2$), respectively. 
We consider three cases corresponding to Sections \ref{sec2}, \ref{sec3} and \ref{sec4}. For all cases we take $\tau_{M,y}=\tau_{M,u}=\tau_{M}$.
In each case, functions ${\bf b}=b_1$, ${\bf c}=c_1$ for $d=1$ and  ${\bf b}=[b_1, b_2]^{\mathrm{T}}$, ${\bf c}=[c_1, c_2]^{\mathrm{T}}$ for $d=2$ are chosen according to Table \ref{gainssKL}, where  \vspace{-0.4cm}
\begin{equation*}
{\scriptsize \begin{array}{ll}
		f_{1}(x)=20x_{1}(x_{2}-x_{2}^{2})\chi_{[0, \frac{a_1}{2} ]\times [0,\frac{a_2}{2}]}(x), \\
		f_{2}(x)=x_{1}(x_{2}-x_{2}^{2})\chi_{[\frac{a_1}{2},\frac{ 3a_1}{4} ]\times [\frac{a_2}{2},a_2]}(x),\\
		f_{3}(x)=(x_1^2-a_1x_1)(x_2^3-a_2x_{2}^2), ~f_{4}(x)=(x_2-a_2)\sin(\frac{2\pi x_1}{a_1}), \\
		f_{5}(x_{1})=\sin(\frac{2x_{1}\pi}{a1})\chi_{[0,\frac{a_1}{2}]}, ~~~~~~~~f_{6}(x_{1})=\sin(\frac{3x_{1}\pi}{a1})\chi_{[\frac{a_1}{3},\frac{2a_1}{3}]},\\
		g_{1}(x)=\chi_{[0, a_1 ]\times [0,\frac{a_2}{2}]}(x), ~~~~~~~~~~
		g_{2}(x)=\chi_{[\frac{a_1}{2}, a_1]\times [0,a_2]}(x),\\
		g_{3}(x_1)=0.2\chi_{[0, \frac{a_1}{4} ]}(x_{1}), ~~~
		~~~~~~~~g_{4}(x_1)=0.2\chi_{[\frac{a_1}{4}, a_{1} ]}(x_{1}). 	\vspace{-0.3cm}
	\end{array} }
\end{equation*}
Here $\chi$ is an indicator function. 
We see that $f_{1},f_{2},g_{1},
 g_{2}\in L^{1}(\Omega)$, $f_{3},f_{4}\in H^{1}(\Omega)$, $f_{3}(x)=f_{4}(x)=0$ for $x\in\Gamma_{D}$, and $g_{3},g_{4},f_{5}, f_{6}\in L^{2}(\Gamma_{N})$. It can be checked that for each case, Assumption \ref{assump1} is satisfied.
 \begin{table}[h]
\center
\caption{Chosen gains $L_{0}$ and $K_{0}$.}
\setlength{\tabcolsep}{2pt}
\scalebox{0.7}{  \begin{tabular}{c|c |c |c c c c c} \hline
 $q=3$, $N_0=1$   & Thm \ref{thm1}  & Thm \ref{thm3}    & Thm \ref{thm2} \\ \hline
$b_1$ & $f_{1}$ & $f_{3}$ & $f_{5}$ \\ 
$c_1$ &  $g_{1}$ & $g_{3}$ &$g_{1}$ \\
$L_{0}$ from \eqref{bL00011} & 1.6349  & 2.1837  & 4.1634\\
$K_{0}$ from \eqref{bL00022} & 1.2696   & 47.3821 & 1.6349 \\ \hline
 $q=8.1$, $N_0=3$             &  Thm \ref{thm1}  & Thm \ref{thm3}    & Thm \ref{thm2}    \\ \hline
$b_1, b_2$ &  $f_{1},f_{2}$  &  $f_{3},f_{4}$ &   $f_{5},f_{6}$  \\
$c_1,c_2$ &  $g_{1}, g_{2}$   & $g_{3}, g_{4}$ &  $g_{1}, g_{2}$  \\ \hline
$L_{0}$ from \eqref{L000} & ${\tiny \left[\begin{array}{ccc}
8.428  &  6.036 \\
   -0.295 &  -0.424 \\
    0.204  &  0.150
\end{array}	 \right]}$  & ${\tiny \left[\begin{array}{ccc}
9.964 &   58.153\\
    0.161 &  -0.416\\
    0.927 &   -0.188
\end{array}	 \right]}$ & ${\tiny \left[\begin{array}{ccc}
7.108 &    4.841 \\
   -0.133 &   -0.525 \\
    0.709  &  0.085
\end{array}	 \right]}$  \\ \hline
 & {\tiny $\delta=0.04$}  & {\tiny $\delta=0.02$}  & {\tiny $\delta=0.05$} \\
$K_0$ from \eqref{designLMI11abc} & ${\tiny \left[\begin{array}{ccc}
5.260 &  0.029 &   -0.034\\
   -0.094 &    0.253 &  -0.097
\end{array}	 \right]}$ &  ${\tiny \left[\begin{array}{ccc}
11.033 &    0.026 &   0 \\
   0 &    0 &   -0.040
\end{array}	 \right]}$  &  ${\tiny \left[\begin{array}{ccc}
7.886 &   -0.280 &    0.385 \\
   -8.444 &    0.039 &    0.518
\end{array}	 \right]}$    \\   \hline
\end{tabular} } \label{gainssKL} 
\end{table}
\vspace{-0.3cm}

For the case that $q=3$ and $N_0=1$, the gains $L_{0}$ and $K_{0}$ are found from \eqref{bL000} with $\delta=1$ and are given in Table \ref{gainssKL}.  The LMIs of Theorems \ref{thm1}, \ref{thm3}, and \ref{thm2} as well as their counterparts by classical Halanay's inequality (Remarks \ref{remark3}, \ref{rem4}, and \ref{rem5}) were verified, respectively, for $N=2,\dots,8$ to obtain maximal values of $\tau_{M}$ ($\delta=\delta_{1}>0$ is chosen optimally) that preserve the feasibility of LMIs. The results are given in Table \ref{observer1unstab}. From Table \ref{observer1unstab}, it is seen that  the vector Halanay inequality always leads to larger delays than the classical scalar Halanay inequality.
\begin{table}[h]
\caption{Max $\tau_{M}$ for feasibility of LMIs ($q=3$, $N_0=1$): Theorems \ref{thm1}, \ref{thm3}, \ref{thm2} (vector Halanay's inequality) VS Remarks \ref{remark3}, \ref{rem4}, \ref{rem5} (classical scalar Halanay's inequality).}  \label{observer1unstab}  
    \centering
    \setlength{\tabcolsep}{2pt}
\scalebox{0.65} {\begin{tabular}{c|cc|cc|cc|cc|cc|cc|cc} \hline
      $N$               & \multicolumn{2}{c|}{ 2 }  & \multicolumn{2}{c|}{  3 }    & \multicolumn{2}{c|}{  4 }  & \multicolumn{2}{c|}{  5 }    & \multicolumn{2}{c|}{  6 } & \multicolumn{2}{c|}{  7 }   & \multicolumn{2}{c}{  8 }      \\ \cline{2-15}
       &  $\delta$  & $\tau_{M}$  & $\delta$   &$\tau_{M}$  &  $\delta$   &$\tau_{M}$ &  $\delta$   &$\tau_{M}$   &  $\delta$   &$\tau_{M}$ &  $\delta$   &$\tau_{M}$ &  $\delta$   &$\tau_{M}$   \\  \hline
Thm \ref{thm1}    & 0.35  & 0.237 & 0.12 & 0.292  & 0.1 & 0.303 & 0.07  & 0.311 & 0.05 & 0.318 & 0.05 & 0.39 &0.04 & 0.323    \\ 
Rmk \ref{remark3} & 1  & 0.196 &  1 & 0.225 & 1 & 0.236 &  0.95 & 0.247 & 0.9 & 0.256 & 0.8 & 0.259  & 0.7 & 0.264   \\       \hline 
Thm \ref{thm3}        & -- & -- & --   & -- & 0.48 & 0.137 & 0.45 & 0.175 & 0.3 & 0.248 & 0.25 & 0.259  & 0.2 & 0.272  \\ 
Rmk \ref{rem4}    & -- & -- & --  & -- & 3 & 0.033 &  2.5 & 0.041 & 1.2 & 0.107 & 1.1 & 0.123 & 1.08 & 0.141  \\ \hline 
Thm \ref{thm2}    & 0.18 & 0.276 & 0.06 & 0.312    & 0.06 & 0.319 & 0.03   & 0.323 & 0.03 & 0.328 & 0.02 & 0.329  & 0.02 &0.331   \\ 
Rmk \ref{rem5}    & 0.9 & 0.222 & 0.8 & 0.257  & 0.6 & 0.266 & 0.6  & 0.275 & 0.5 & 0.281 &  0.4 & 0.285 &  0.3 &0.291    \\ \hline 
\end{tabular} }    
\end{table} 
\vspace{-0.3cm}

For the case that $q=8.1$ and $N_0=3$, we found that the $L_{0}$ and $K_{0}$ obtained from \eqref{bL000} were not efficient for the feasibility of LMIs of Theorems \ref{thm1}, \ref{thm3}, \ref{thm2} and Remarks \ref{remark3}, \ref{rem4}, \ref{rem5} even for $\tau_{M,y}=\tau_{M,u}=0$. 
We design $L_{0}$ ($\delta=\delta_1=0.01$, $N=20$) and $K_{0}$ ($N=30$) from \eqref{L000} and \eqref{designLMI11abc} in Remark \ref{L0K0design} and give the values in Table \ref{gainssKL}. The LMIs of Theorems \ref{thm1}, \ref{thm3}, and \ref{thm2} as well as their counterparts by classical Halanay's inequality (Remarks \ref{remark3}, \ref{rem4}, and \ref{rem5}) were verified, respectively, for different $N$ to obtain maximal values of $\tau_{M}$ ($\delta=\delta_{1}>0$ is chosen optimally) that preserve the feasibility of LMIs. 
The results are given in Table \ref{observer}. From Table \ref{observer}, it is seen that the vector Halanay inequality leads to larger delays than the classical scalar one for comparatively large $N$, whereas for comparatively small $N$, the classical scalar Halanay inequality leads to larger delays. This phenomenon corresponds to Remark \ref{remkdelta1}. 
\vspace{-0.2cm}
\begin{table}[h]
\caption{Max $\tau_{M}$ for feasibility of LMIs ($q=8.1$, $N_0=3$): Theorems \ref{thm1}, \ref{thm3}, \ref{thm2} (Vector Halanay's inequality) VS Remarks \ref{remark3}, \ref{rem4}, \ref{rem5} (Classical Scalar Halanay's inequality).}  \label{observer}  
    \centering
    \setlength{\tabcolsep}{2pt}
\scalebox{0.65} {\begin{tabular}{c|cc|cc|cc|cc|cc|cc} \hline
      $N$               & \multicolumn{2}{c|}{ 20 }  & \multicolumn{2}{c|}{  $25$ }    & \multicolumn{2}{c|}{  $30$ }  & \multicolumn{2}{c|}{  $35$ }    & \multicolumn{2}{c|}{  $40$ } & \multicolumn{2}{c}{  $45$ }        \\ \cline{2-13}
       &  $\delta$  & $\tau_{M}$  & $\delta$   &$\tau_{M}$  &  $\delta$   &$\tau_{M}$ &  $\delta$   &$\tau_{M}$   &  $\delta$   &$\tau_{M}$ &  $\delta$   &$\tau_{M}$   \\  \hline
Thm \ref{thm1}    & 0.051  & 0.0104 & 0.049 & 0.0342  & 0.048 & 0.0414 & 0.047  & 0.0454 & 0.045 & 0.0481 & 0.045 & 0.0502     \\ 
Rmk \ref{remark3} & 4.5  & 0.0267 &  4 & 0.0330 & 3 & 0.0357 &  3 & 0.0376 & 2.8 & 0.0395 & 2.5 & 0.0407     \\       \hline 
 $N$               & \multicolumn{2}{c|}{  $30$ }  & \multicolumn{2}{c|}{  $35$ }    & \multicolumn{2}{c|}{  $40$ } & \multicolumn{2}{c|}{  $45$ }    & \multicolumn{2}{c|}{  $50$ }  & \multicolumn{2}{c}{  $55$ }       \\ \cline{2-13}
  &  $\delta_{{\mathrm{opt}}}$  & $\tau_{M}$  & $\delta_{{\mathrm{opt}}}$   &$\tau_{M}$  &  $\delta_{{\mathrm{opt}}}$   &$\tau_{M}$ &  $\delta$   &$\tau_{M}$   &  $\delta$   &$\tau_{M}$ &  $\delta$   &$\tau_{M}$   \\  \hline
Thm \ref{thm3}        & 0.019 & 0.0168 & 0.018   & 0.0219 & 0.018 & 0.0271 & 0.017 & 0.0301 & 0.017 & 0.0311 & 0.017 & 0.0337    \\ 
Rmk \ref{rem4}    & 6 & 0.0206 & 6  & 0.0215 & 5 & 0.0230 &  4.5 & 0.0238 & 4 & 0.0240 & 4 & 0.0254   \\ \hline 
   $N$               & \multicolumn{2}{c|}{ 7 }  & \multicolumn{2}{c|}{  8 }    & \multicolumn{2}{c|}{ 9 }  & \multicolumn{2}{c|}{  10 }    & \multicolumn{2}{c|}{  15 } & \multicolumn{2}{c}{  20 }         \\ \cline{2-13} 
    &  $\delta_{{\mathrm{opt}}}$  & $\tau_{M}$  & $\delta_{{\mathrm{opt}}}$   &$\tau_{M}$  &  $\delta_{{\mathrm{opt}}}$   &$\tau_{M}$ &  $\delta$   &$\tau_{M}$   &  $\delta$   &$\tau_{M}$ &  $\delta$   &$\tau_{M}$   \\  \hline
Thm \ref{thm2}    & -- & -- & 0.15 & 0.0112    & 0.15 & 0.0242 & 0.15   & 0.0291 & 0.14 & 0.0467 & 0.12 & 0.0506    \\ 
Rmk \ref{rem5}    & 7 & 0.0036 & 6 & 0.0106  & 5 & 0.0136 & 5  & 0.0151 & 2.5 & 0.0254 &  2 & 0.0311     \\ \hline 
\end{tabular} }    
\end{table}

\vspace{-0.4cm}

\begin{figure}[h]
\centerline{\includegraphics[width=7cm]{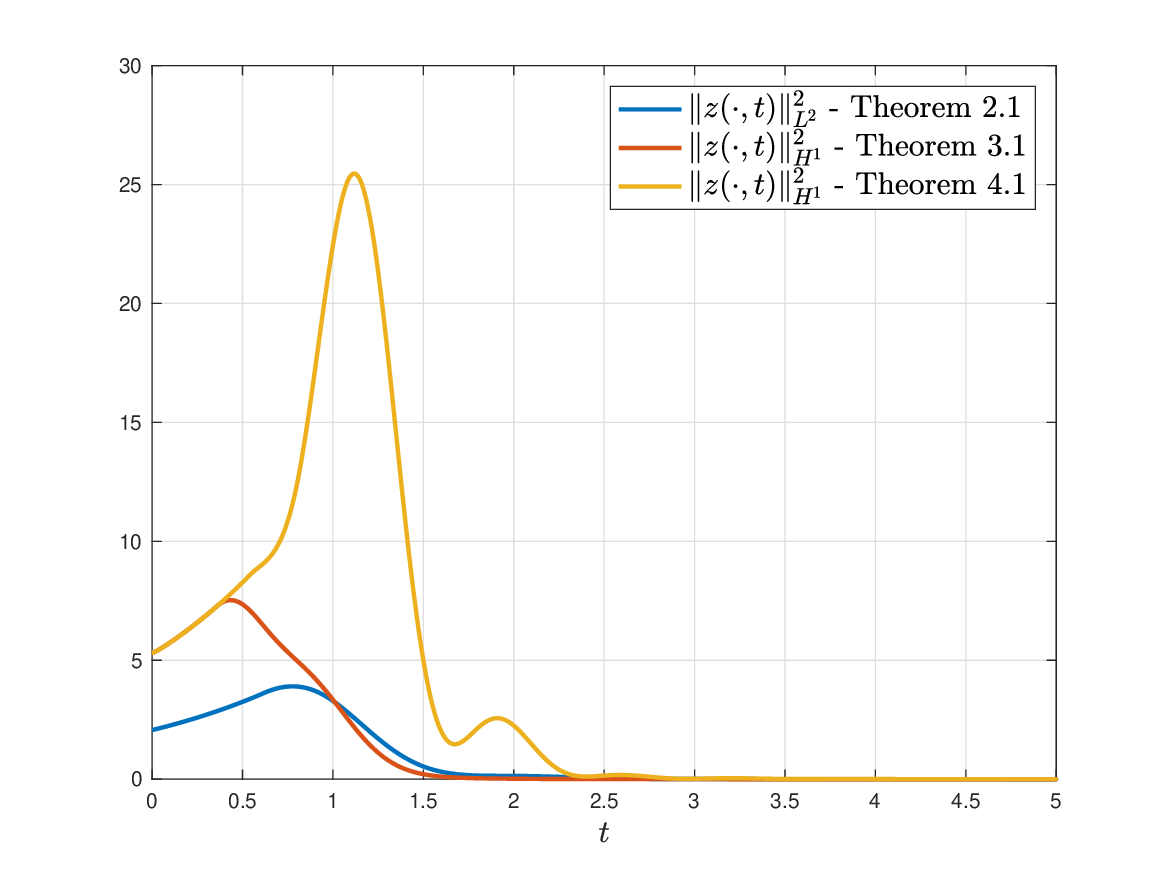}} \vspace{-0.25cm}
\caption{Evolutions $\|z(\cdot,t)\|^{2}_{L^{2}}$ (Theorem \ref{thm1}), $\|\nabla z(\cdot,t)\|^{2}_{L^{2}}$ (Theorem \ref{thm3}), and $\|z(\cdot,t)\|^{2}_{L^{2}}$ (Theorem \ref{thm2}) VS $t$.}
\label{fig1}
\end{figure}

For simulation of closed-loop systems studied in Sections \ref{sec2}, \ref{sec3} and \ref{sec4}, we consider the case $q=3$, $N_0=1$  and fix $N=5$.  Consider time-varying delays $\tau_{y}(t)=\frac{\tau_{M}}{2}[1+\sin^2 t]$ and $\tau_{u}(t)=\frac{\tau_{M}}{2}[1+\cos^2 t]$ (corresponding maximal values of $\tau_{M}$ are chosen as 0.311, 0.175, and 0.323, respectively according to Table \ref{observer}). We approximate the solution norm using 150 modes as $\|z(\cdot,t)\|^{2}_{L^{2}}\approx \sum_{n=1}^{150}z_{n}^{2}(t)$ and $\|\nabla z(\cdot,t)\|^{2}_{L^{2}}\approx \sum_{n=1}^{150}\lambda_{n}z_{n}^{2}(t)$. 
Take initial conditions $z_{0}(x)=x_{1}(a_1-x_{1})\cos(\frac{\pi}{2a_2}x_{2})$. The closed-loop systems (with the tail ODEs truncated after 150 modes) are simulated using MATLAB. The simulations are presented in Fig. \ref{fig1}. The numerical simulations validate the theoretical results.
Stability of the closed-loop systems in simulations was preserved for $\tau_{M}= 0.48$ for Theorem \ref{thm1}, $\tau_{M}=0.38$ for Theorem \ref{thm3}, and $\tau_{M}=0.42$ for Theorem \ref{thm2}, which may indicate that our approach is somewhat conservative in this example.\vspace{-0.3cm}

\vspace{-0.2cm}
\section{Conclusions}\label{sec5} \vspace{-0.4cm}
We considered the finite-dimensional observer-based control of 2D linear heat equation with fast-varying unknown input and known output delays. To compensate the output delay that appears in the infinite-dimensional part of the closed-loop system, we suggested a vector Lyapunov functional combined with vector Halanay's inequality. 
In the numerical examples, the vector Halanay inequality led to larger delays for larger dimensions of the observer that preserve the stability than the classical one. 
Improvements and extension of the results to various high-dimensional PDEs may be topics for future research.

\bibliographystyle{abbrv}        
\bibliography{ref}           
\end{document}